\documentclass[preprint,12pt]{elsarticle}

\usepackage{amssymb}
\usepackage{graphicx}
\usepackage{amsmath}
\usepackage{xcolor}
\usepackage[T1]{fontenc}

\journal{Physics of the Dark Universe}

\begin{document}

\begin{frontmatter}



\title{Hermeian haloes: extreme objects with two interactions in the past}


\author[inst1,inst2]{Anastasiia Osipova}

\affiliation[inst1]{organization={National Research University Higher School of Economics, Faculty of Physics},%Department and Organization
            addressline={Myasnitskaya 20}, 
            city={Moscow},
            postcode={101000}, 
            country={Russia}}
\affiliation[inst2]{organization={P.N. Lebedev Physical Institute of the Russian Academy of Sciences},%Department and Organization
            addressline={Profsojuznaja 84/32 Moscow}, 
            city={Moscow},
            postcode={117997}, 
            country={Russia}}

\author[inst2]{Sergey Pilipenko}

\author[inst3]{Stefan Gottl\"{o}ber}
\author[inst3]{Noam I. Libeskind}
\author[inst4]{Oliver Newton}
\author[inst8,inst5,inst3]{Jenny G. Sorce}
\author[inst6,inst7]{Gustavo Yepes}

\affiliation[inst3]{organization={Leibniz-Institut f\"{u}r Astrophysik (AIP)},%Department and Organization
            addressline={An der Sternwarte 16}, 
            city={Potsdam},
            postcode={D-14482}, 
            country={Germany}}
\affiliation[inst4]{organization={Center for Theoretical Physics of the Polish Academy of Sciences},%Department and Organization
            addressline={Al. Lotnik\'{o}w 32/46}, 
            city={Warsaw},
            postcode={02-6682}, 
            country={Poland}}
\affiliation[inst8]{organization={Univ. Lille, CNRS},%Department and Organization
            addressline={Centrale Lille, UMR 9189 CRIStAL}, 
            city={Lille},
            postcode={F-59000}, 
            country={France}}
\affiliation[inst5]{organization={Universit\'e Paris-Saclay, CNRS, Institut d'Astrophysique Spatiale},%Department and Organization
            addressline={}, 
            city={Orsay},
            postcode={91405}, 
            country={France}}
\affiliation[inst6]{organization={Departamento de F\'{i}sica Te\'{o}rica, M\'{o}dulo 15, Facultad de Ciencias, Universidad Aut\'{o}noma de Madrid},%Department and Organization
            addressline={}, 
            city={Madrid},
            postcode={28049}, 
            country={Spain}}
\affiliation[inst7]{organization={Centro de Investigaci\'{o}n Avanzada en F\'{i}sica Fundamental (CIAFF), Facultad de Ciencias, Universidad Aut\'{o}noma de Madrid},%Department and Organization
            addressline={}, 
            city={Madrid},
            postcode={28049}, 
            country={Spain}}

\begin{abstract}
Recent studies based on numerical models of the Local Group predict the existence of field haloes and galaxies that have visited both the Milky Way and M31 in the past, called Hermeian haloes. We extend this analysis beyond the Local Group using two high-resolution dark matter-only N-body simulations from the MultiDark suite. We define Hermeian haloes as field haloes which had close interactions with two other more massive field haloes in the past, called targets. We find that Hermeian haloes are a more extreme example of field haloes with interactions in the past than the well-known backsplash haloes that experienced only one interaction. Compared to backsplashers, Hermeians have more concentrated density profiles and tend to occupy more overdense regions. They also have higher velocities relative to their target haloes and relative to their neighbours within 1~$h^{-1}\; \mathrm{Mpc}$. Hermeian haloes can be found around every halo in the simulation (if the resolution is sufficient) and make up 0.4 to 2.3 per cent of the total number of field haloes (for haloes more massive than $10^{10}\; h^{-1}\;\mathrm{M_{\odot}}$ and $3.3 \times 10^{7}\; h^{-1}\;\mathrm{M_{\odot}}$, respectively), increasing to 10 per cent in overdense regions. They tend to be distributed close to the line connecting their targets, which may help to identify Hermeian haloes in observations. We also identify Local Group analogues in the simulation and find that about one-third (15 out of 49) of them contain Hermeian haloes if the distance between the two main haloes is below 1~$h^{-1} \;\mathrm{Mpc}$.

\end{abstract}



\begin{keyword}
large scale structure \sep halo interaction \sep cosmological simulation
%% keywords here, in the form: keyword \sep keyword \sep 
\PACS 0000 \sep 1111
\MSC 0000 \sep 1111
\end{keyword}

\end{frontmatter}

\section{Introduction}
The hierarchical paradigm of galaxy formation predicts that galaxies experience a lot of interactions with other galaxies on cosmological timescales. The most common interactions are galaxy mergers, but recent studies based on cosmological simulations have revealed a more complex picture with a variety of halo evolution paths. These include `backsplash' haloes (sometimes called `flyby' haloes) that fell into a halo (called a target halo) at early times and then left it \cite{Gill:Knebe:Gibson:2005,Moore04,Knebe11a,An:Kim:Moon:2019}. Studies have shown that around 10 per cent of field haloes in the Local Group (LG) can be backsplashers \cite{Ludlow09,Teyssier:Johnston:Kuhlen:2012,GarrisonKimmel14,Bakels21,Green21}. %If such objects host galaxies, they will have properties that differ from field galaxies that have not experienced interactions with the \textbf{more massive object} in the past \cite{Balogh00,Knebe11a,Simpson18,Buck19,Benavides21,Joshi21}. \textbf{The most obvious consequences of flying through a larger halo are the removal of gas and quenching of the star formation.} 
Simulations of the LG have also revealed a class of `renegade' haloes that pass through one of the LG main galaxies and are then accreted into the other \cite{Knebe11b,Wetzel15}. %Renegade haloes differ from the rest of the subhalo population by their anisotropic spatial distribution. 
Most of these results come from constrained simulations of the LG. Unlike typical simulations, constrained simulations stem from a set of constraints that can be either redshift surveys or radial peculiar velocities. Analogues of backsplash and renegade haloes beyond the LG-like systems, as well as other types of halo evolution channels have been also analysed \cite{vandenBosch17}.

Hydrodynamical simulations show that few, if any, stars are stripped from low-mass galaxies when they interact, so the overall luminosity of a galaxy flying through another galaxy should not change \cite{Knebe11a}. On the other hand, the gas is removed which quenches the star formation \cite{Balogh00,Simpson18,Buck19,Benavides21,Joshi21}. In particular, 100\% of red field Ultra-Diffuse Galaxies (UDGs) and 70 to 100\% of the quenched field dwarfs may be backsplash galaxies \cite{Benavides21,Joshi21}. Hence, backsplash galaxies become an important tool for the study of quenching \cite{Casey22}.

A recent study \cite{Newton} described a new class of field haloes that passed through the haloes of both the Milky Way (MW) and the Andromeda galaxy (M31) -- the so-called Hermeian haloes (named after Hermes, the Greek god of travellers). The authors have analyzed three high resolution zoom-in constrained simulations of the LG from the HESTIA \cite{Libeskind2020} suite.  They found more than one hundred such haloes in them. Hermeian haloes have several distinctive properties that make them interesting to study \cite{Newton}. First, two episodes of interaction significantly increase the dark matter density in the centers of Hermeian haloes. This makes Hermeian haloes interesting for indirect searches of dark matter particles, since higher density leads to higher frequency of possible dark matter particle annihilation \cite{Gaskins:2016}. Second, they are clustered along the line connecting the MW and M31, and if they host galaxies, they will have chemical signatures of their past interactions. This could provide a method to distinguish Hermeian galaxies from regular field dwarfs in observations of the real LG. Also, like backsplash haloes, most of the Hermeians are receding from their second targets, so they may offer a possible explanation for galaxies with `strange' peculiar velocities, such as NGC 3109 \cite{Banic:Zhao:2017,Banik:Haslbauer:Pawlowski:2021}. Hermeian galaxies in the simulations analyzed by \cite{Newton} facilitate the exchange of gas and stars between the MW and M31. This indicates that such objects may pollute the gas in galaxies with chemical elements made in other galaxies. However, the limited statistics available in the first study of Hermeian haloes \cite{Newton} leaves many open questions about properties and origin of Hermeian haloes.
 
The primary aim of this paper is to further study this new class of field haloes, Hermeian haloes. We answer questions regarding how common these haloes are in the Universe and across analogues of the Local Group and how much they differ from regular field haloes and backsplash haloes. For this purpose, we first clarify and generalize the definition of these objects: in \cite{Newton} Hermeians were related to the Milky Way and M31, but we analyze other kinds of galaxy pairs, even those that bear minimal resemblance to the LG. %I would omit this since it is not touched upon in this paper: Hermeian haloes in the Universe may be responsible for matter exchange between galaxies, and, in particular, the exchange of life precursors, which is important for the hypothesis of panspermia.%
%It is also interesting to find out how frequent Hermeians are in the Universe and in which systems they occur more often. 
Thus, we apply our definition of Hermeian halo to two high resolution cosmological dark matter-only simulations. We replicate the study of dark matter density and spatial distribution of Hermeian haloes that was carried out in \cite{Newton} using a richer sample of tens of LG analogues, which provides a significant improvement in statistical power. We do not use constrained realizations of the Local Group, but several LG-like objects can be selected from our simulations. The properties of the LG analogues we select vary considerably (for example, the MW-M31 distance), so we analyse which properties determine the presence and abundance of Hermeian haloes. Last but not least, we study the effect of numerical errors on the halo sample and develop an algorithm for reliably detecting Hermeian haloes in simulations.

One may ask why a dark matter-only simulation is sufficient for the goals of the current study. In the previous work on Hermeians \cite{Newton} there were only 4 Hermeian galaxies, which retain stars and gas, but they have the same trajectories as the dark Hermeian haloes (i.e. those without a baryonic component). Therefore, to study the spatial distribution of Hermeian galaxies and develop criteria for their identification in observations, one may use dark matter-only simulations, at least as a first approximation. 

%\hfill \break
This paper is structured as follows: in Section~\ref{sec:Simulations and Methodology} we describe the cosmological simulations analyzed in this work and the method we used to identify haloes passing through another halo. In this section we also introduce our definitions the halo classes we study. In Section~\ref{sec: Results} we summarize our main results and discuss their consequences. Finally, Section~\ref{sec: Conclusions} draws the conclusion and identifies areas for further research.

\section{Simulations and Methodology}\label{sec:Simulations and Methodology}

%In this section we briefly review the simulations used throughout this study, describe our algorithm for finding haloes that have passed through another halo, and define the analyzed halo classes. 
\subsection{Simulations and halo catalogues}
The previous paper on Hermeian haloes \cite{Newton} shows that a high mass resolution is critical to resolve Hermeian haloes in LG-like systems: most of these haloes are low mass, below $10^9$~M$_\odot$. The simuations in \cite{Newton} had dark matter particle mass $m_{DM}=1.5\times 10^5$~$h^{-1}\;\mathrm{M_{\odot}}$ which is sufficient to resolve dwarf-galaxy-sized haloes to hundreds of particles.
We use two dark matter-only $N$-body simulations: Very Small MultiDark Planck (VSMDPL) and Extremely Small MultiDark Planck (ESMDPL) carried out by an international consortium within the framework of the MultiDark project\footnote{http://www.multidark.es}. These simulations have highest mass resolution among the MultiDark suite, in particular, ESMDPL has particle mass sufficient to resolve $M=3\times10^7$~M$_\odot$ haloes. 
Both simulations were performed with the GADGET-2 code \cite{Springel} and used Planck Cosmology ($h = 0.6777$, $\Omega_{\mathrm{\Lambda}} = 0.6929$, $\Omega_{\mathrm{m}} =  0.3071$, $\Omega_{\mathrm{b}} = 0.0482$, $\sigma_{\mathrm{8}} = 0.8228$, $n_{\mathrm{s}} = 0.96$ \cite{Planck2015}). The parameters of simulations are summarized in Table~\ref{tab:table_1}. For ESMDPL we have 68 snapshots between z = 28 and z = 0 with a time spacing varying from 0.009 to 0.7 Gyr and for VSMDPL we use 22 snapshots between z = 0.6 and z = 0 with a time spacing varying from 0.2 to 0.4 Gyr. %I think you should omit this sentence because above youi explicitely say your not using constrained sims and its just confusing. That the ESMDPL is constrained is irrelevant for this paper:%
The ESMDPL simulation was set up using a "constrained realization" of the initial conditions. When evolved to z = 0, it reproduces the observed large-scale environment surrounding the LG, and contains analogues of the Virgo Cluster and the LG. That ESMDPL is a constrained simulation is unimportant for our analysis; however, as with any constrained simulation it can also be used for statistics of objects in the given volume. %\textbf{ A complete description of the simulations is avaliable in ...???}
% Numerical parameters table
\begin{table}
	\centering
	\caption{Numerical parameters for the simulations used in this paper. The columns give the simulation name, the size of the simulation box in $h^{-1} \;\mathrm{Mpc}$, the number of particles, the mass per simulation particle $m_{\mathrm{p}}$ in units $h^{-1}\;\mathrm{M_{\odot}}$, the total number of simulation outputs stored, $N_{\mathrm{out}}$, the redshift of the initial conditions $z_{\mathrm{init}}$, the Plummer equivalent gravitational softening length $\epsilon$ at the start of the simulation ($\epsilon_{z}$) and at low redshift ($\epsilon_{0}$) in $h^{-1} \;\mathrm{kpc.}$}
	\label{tab:table_1}
	\begin{tabular}{lccccccr} % four columns, alignment for each
		\hline
		Simulation & box & particles & $m_{\mathrm{p}}$ & {$N_{\mathrm{out}}$} & $z_{\mathrm{init}}$ & $\epsilon_{0}$ & $\epsilon_{z}$\\
		\hline
		VSMDPL & 160 & $3840^{3}$ & $6.2 \times 10^{6}$ & 150 & 150 & 2 & 1\\
		ESMDPL & 64 & $4096^{3}$ & $3.3 \times 10^{5}$ & 70 & 150 &1.0 & 0.5 \\
		\hline
	\end{tabular}
\end{table}

The results reported here use halo data stored in catalogues created using the Consistent Trees \cite{Behroozi:Wechsler:Wu:2013:763:18} and Robust Overdensity Calculation using K-Space Topologically Adaptive Refinement algorithms (ROCKSTAR; \cite{Behroozi:Wechsler:Wu:762:2}). ROCKSTAR catalogues are designed to provide a smooth history of each halo, so the properties of the halo can be interpolated if it fails to be identified in some intermediate snapshot (due to e.g. resolution). Haloes are traced in time by  ROCKSTAR with high accuracy:  the fraction of haloes that do not have physically consistent progenitors is between 1 and 2 per cent (see \cite{Behroozi:Wechsler:Wu:2013:763:18}, for more details).

Throughout this paper we use the virial radius, $\mathrm{R_{vir}}$, defined by ROCKSTAR in terms of the virial spherical overdensity with respect to the mean background density \cite{Behroozi:Wechsler:Wu:2013:763:18, Bryan:Norman:1998}. For our cosmological parameters, at $z = 0$ $\mathrm{R_{vir}}$ is defined as the radius of a sphere with overdensity of $\sim 326$ of the averaged matter density.

Due to the large number of VSMDPL snapshots, the structure that stores information about Hermeian haloes and their targets does not fit into the memory of the computer we used. Consequently, only the snapshots that correspond to redshift, $z \leq 0.61$ were analyzed. For ESMDPL, the full set of available snapshots was analysed. Therefore, the VSMDPL simulation is used only to show the convergence of the results.

\subsection{Passage detection algorithm}
\label{sec:passage}
In order to identify Hermeian and backsplash haloes, as well as their targets, we first identify every passage through another halo for each halo during its evolution. In this Subsection we describe the search for passages, and in the next Subsection we give our definitions of Hermeian and backsplash haloes.
The ROCKSTAR and Consistent Trees generate catalogues of haloes for each snapshot, as well as their commonly used characteristics (see appendices of \cite{Rodrigues-Puebla} for an overview of the halo catalogues). The proposed catalogues store more than 80 halo characteristics, and the properties that are used by our algorithm are reviewed below.

\texttt{ID:} The halo \texttt{ID} that is guaranteed to be unique across all snapshots of the entire simulation.

\texttt{Desc\_id:} The halo \texttt{ID} of the descendant halo. A halo at one time step (a progenitor) is linked to a halo at the next time step (the descendant) if the majority of the particles in the progenitor end up in the descendant.

\texttt{Pid:} The parent halo \texttt{ID}. For field haloes (those that are not subhaloes) this is -1. Otherwise, it is the halo \texttt{ID} of the smallest host halo that contains this halo.

\texttt{Desc\_pid:} The \texttt{Pid} of the descendant halo.

\hfill \break
Using \texttt{ID} and \texttt{Desc\_id}, we track halo descendants throughout the simulation, and using \texttt{pid} and \texttt{Desc\_pid} we register the passage of the halo through another halo. 
In each simulation snapshot, we select haloes whose $\texttt{pid}~\neq-1$ and $\texttt{desc\_pid} = -1$. These conditions mean that at the considered snapshot a halo is inside a host halo and at the next one the halo (or more precisely, its descendant) is a field halo. Therefore, we classify such objects as having passed through a second halo. Based on the analysis of the DM halo mass function recovered by the ROCKSTAR halo finder in \cite{Knebe:Knollman:Muldrew:2011}, we consider only haloes having at least $N_p=24$ particles at the snapshot when the passage is identified.
At the same time, we record the history of changes in two properties (coordinates and $\mathrm{M_{vir}}$) of the descendants of previously selected haloes from the moment of passage through another halo to z = 0. We also track descendants of haloes that have been passed through (i.e. descendants of targets).

In this way, we checked all haloes in the simulation at each time step and selected those haloes that performed exactly one or exactly two passages. From these two groups, we selected for further consideration only those haloes whose virial mass did not increase by more than a factor of two between any successive snapshots. This ensures that in most cases we look at the main branch of the merger tree. This is an important check, because if we restrict ourselves only to the criterion for \texttt{pid} and \texttt{desc\_pid} without checking the mass of the haloes, we will track the haloes with which backsplash or Hermeian candidates merge and not the backsplash and Hermeian haloes themselves. 

%The next section gives the definition of selected haloes and their classification into backsplash and Hermeian haloes.

\subsection{Definitions of halo classes}
\label{section: Definitions of halo classes}
There is a degree of uncertainty around the terminology in the area of haloes passing through another halo: backsplash haloes are often called ``flyby''. Moreover, the definitions of these haloes vary in the literature \cite{Gill:Knebe:Gibson:2005, Teyssier:Johnston:Kuhlen:2012, Diemer_2021, Haggar:Gray:Pearce:2020, Mansfield:Kravtsov:2020}. 

\hfill \break
In this paper, to be called backsplash, a halo must satisfy the following conditions:
\begin{flushleft} 
\begin{enumerate}
    \item a backsplash halo is a field halo at $z = 0$, %the descendant of the halo which passed through a target at some time ago must be a field halo (i.e. $\textbf{pid} = -1$). % the descendant of its progenitor?
    \item its main progenitor was within the virial radius of another more massive halo, the target, (i.e. it had $\texttt{pid}~\neq-1$, $\texttt{desc\_pid} = -1$ and $N_p\geq 24$) only once during the time when the halo finder is able to track its evolution (this time is different for each halo),
    \item a backsplash halo and the descendant of the target are different haloes at $z = 0$.
    
\end{enumerate}
\end{flushleft}
\hfill \break
In this definition it is important to note that we consider haloes that experienced only one passage through the target. Thus, Hermeian haloes are not backsplashers, which makes it more convenient to compare backsplash and Hermeian samples. An alternative choice is to consider haloes that had one or more passages, which is adopted in e.g. \cite{Diemer_2021}.

To be called Hermeian, a halo must satisfy the following conditions:
\begin{flushleft} 
\begin{enumerate}
    \item a Hermeian halo is a field halo at $z = 0$,
    \item its main progenitor was within the virial radius of two other haloes (the targets) exactly once each (during the time when the halo finder is able to track its evolution),
    \item the descendants of the targets are different at $z = 0$,
    \item a Hermeian halo and the descendants of the targets are different haloes at $z = 0$.
     
\end{enumerate}
\end{flushleft}
\hfill \break
In both cases, the descendants of the targets of Hermeians and backsplashers may be subhaloes at z = 0.
\hfill \break
This definition of Hermeian haloes closely resembles the one in \cite{Newton}. But it can be modified according to the purposes of a study. For example, for the exploration of matter exchange one should not restrict the analysis to field haloes at $z = 0$. As we mentioned in the Introduction, renegade haloes, which passed through one halo and become a subhalo in another one, could be quite common in the LG.

We don't include haloes with three or more passages in our definition of a Hermeian because this choice simplifies subsequent analysis. Distances between targets, target mass rations, etc, can be computed uniquely when every Hermeian has only two targets.
%Accordingly, in this work backsplash haloes are field haloes (at $z = 0$) that have passed through another halo only once, and Hermeian haloes are field haloes (at $z = 0$) that have passed through the two different haloes. 
%\textcyrillic{Написать про то, что можно вообразить еще множество других не менее интересных классов гало. И про то, что "определение" зависит от целй исследования}

We also compare the properties of Hermeian haloes with those of double-backsplash haloes, which we define as follows:
\begin{flushleft} 
\begin{enumerate}
    \item a double-backsplash halo is a field halo at $z = 0$, %the descendant of the halo which passed through a target at some time ago must be a field halo (i.e. $\textbf{pid} = -1$). % the descendant of its progenitor?
    \item its main progenitor was within the virial radius of another more massive halo (the target) exactly twice during the time when the halo finder is able to track its evolution,
    \item a double-backsplash halo and the descendant of the target are different haloes at $z = 0$.
    
\end{enumerate}
\end{flushleft}
\hfill \break
\section{Results} \label{sec: Results}
\subsection{Numerical effects}
\label{section: Numerical effects}
Halo-finding algorithms are used to identify dark matter haloes in the distribution of simulation particles. %It has been demonstrated that halo finders can recover characteristics of objects consisting of a small number of particles, namely, less than 24, incorrectly \cite{Knebe:Knollman:Muldrew:2011}.%
It is well known that all halo finders  miss or incorrectly  identify haloes with a small number of particles. This critical number depends on the halo finder used (see for a discussion \cite{Knebe:Knollman:Muldrew:2011}). In the following, we use the ROCKSTAR halo finder and start with a minimum halo size of 24 particles. 
As a result, when investigating the Hermeian haloes, we consider only those objects which contain more than 24 particles at the moment of the first passage through another halo (i.e. the last shapshot where it is inside the virial radius of the first target). Although several studies have shown that ROCKSTAR is able to identify subhaloes with 20 particles properly, in some cases it may lose track of haloes with fewer than approximately 30 particles \cite{Knebe:Knollman:Muldrew:2011}. In Section~\ref{section: simulations with different resolution} we discuss the limitations due to mass resolution and determine the limit to which we can trust our analysis. For a detailed study of various halo-finding algorithms abilities to track haloes and to recover their properties, please see \cite{Knebe:Knollman:Muldrew:2011}.

The ability of ROCKSTAR to track precisely backsplash or Hermeian haloes and determine their characteristics has not been explored before. Moreover, the specific properties of Hermeian haloes (e.g. passages through another halo, which are accompanied by the destruction of external parts of the passing halo as a result of tidal interactions), could make them more challenging to track than field haloes with no such interactions in the past and subhaloes. For this reason, we carry out a preparatory study to compare the performance of ROCKSTAR in two simulations with different mass resolutions. We also compare the fraction of backsplash haloes with a previous study made in the literature.
%\begin{figure}
%\centering\includegraphics[width=\linewidth]{Mi_Mz0_relation.pdf}
%\caption{Distribution of the mass ratio of backsplash and Hermeian haloes at z = 0 and at the time of the first passage. }
%\label{fig: Mi_Mz0_relation}
%\end{figure}
We note that the mass of the progenitor of a backsplash or Hermeian halo given by ROCKSTAR at the snapshot when it is first detected by our algorithm (i.e. the last snapshot before it leaves the virial radius of the first target) is, on average, higher than the mass at $z=0$. For this reason we only discuss the impact on the results of the minimum number of particles in the halo at $z=0$ , and not its effect at earlier redshifts.
%The mass of haloes, on average, increases with time \cite{McBride:Fakhouri:Ma:2009}, however, we have found out that most of the backsplashers and Hermeians lose some of their mass by z = 0. Figure~\ref{fig: Mi_Mz0_relation} illustrates this point clearly: the median value of the mass ratio for these haloes at z = 0 and at the time of the first passage through the target is  $\sim 0.8$ and $\sim 0.6$ respectively. For this reason, we determine the acceptable minimum number of particles in a halo based on its mass at z = 0.

\subsubsection{Simulations with different mass resolution}
\label{section: simulations with different resolution}
At fixed halo mass, a halo from the ESMDPL simulation will contain more particles than a halo from the VSMDPL simulation because ESMDPL has a higher mass resolution. Thus since  numerical effects depend not on the mass, but on the number of particles within a halo, we check the results of VSMDPL against those obtained from ESMDPL. In this case, at fixed halo mass the fractions of backsplash haloes in the simulations should show the same trend, and any significant discrepancies would be the result of errors in the operation of the halo-finding algorithm. The two simulations we use have different box sizes, different realizations of the initial perturbation fields and different snapshot frequency, but these differences cannot result in an abrupt change of the number of backsplash haloes at some particular mass, we expect them to affect only the full number of backsplashers.

\begin{figure*}
\centering\includegraphics[width=\textwidth]{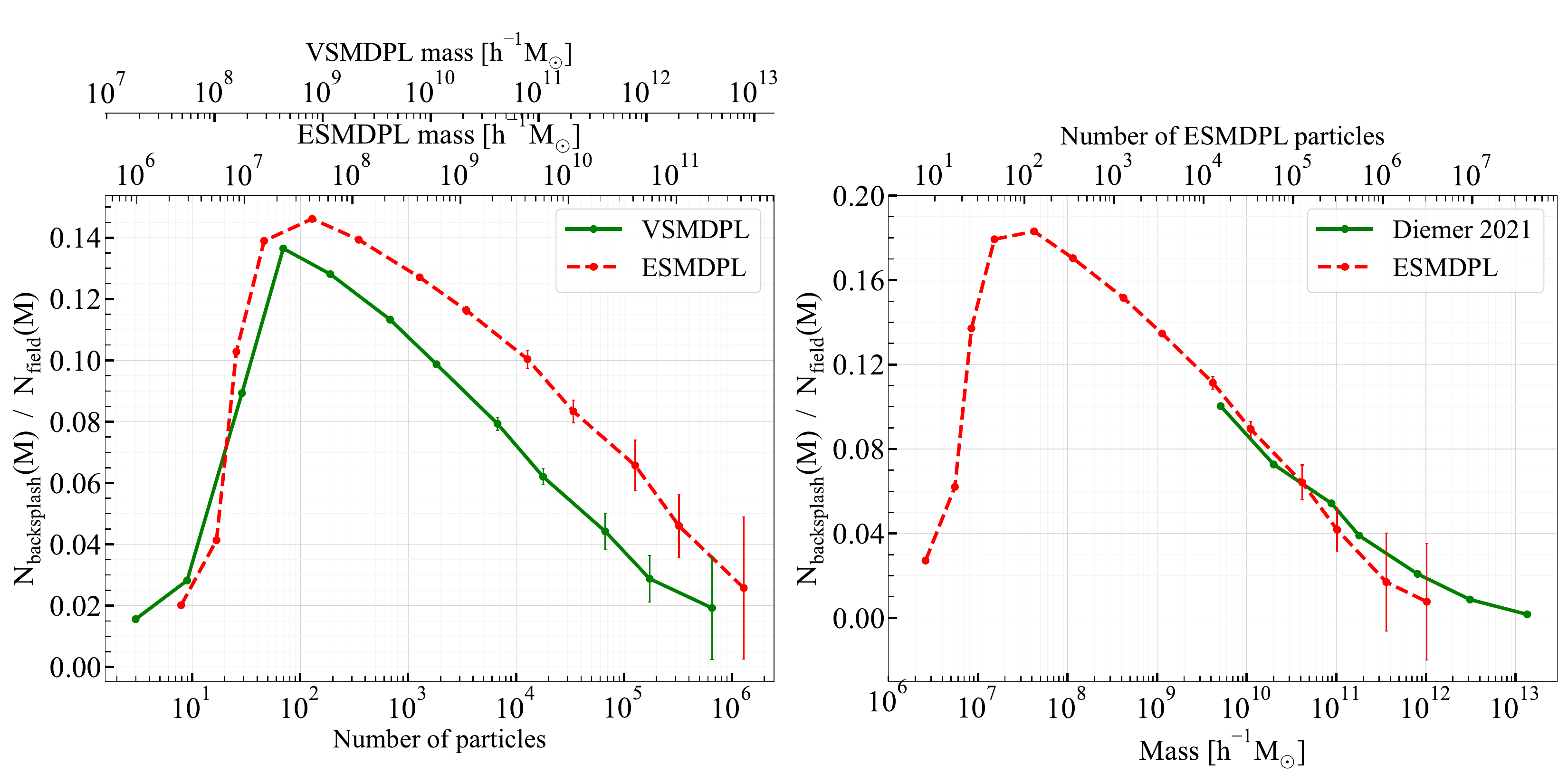}
\caption{\textit{Left panel:} Comparison of the fractions of backsplash haloes obtained with the VSMDPL (solid line) and ESMDPL (dashed line) simulations as a function of number of particles. We use the backsplash haloes that have performed a passage only within the last 6 Gyr. Each point corresponds to the number of backsplash haloes normalised by the total number of field haloes in a mass bin. \textit{Right panel:} Comparison of the backsplash halo fraction in the ESMDPL (dashed line) with the results from the \cite{Diemer_2021} (solid line) as a function of halo mass. We plot the curve from fig.4 of \cite{Diemer_2021} that corresponds to the bound-only mass $M_{\mathrm{200m,bnd}}$ and bound-only virial radius $R_{\mathrm{vir,bnd}}$).  In this figure we adopt the \cite{Diemer_2021} definition of backsplash haloes as objects that made one or more passages. In the rest of this paper we adopt the definition that backsplash haloes experienced only one passage through a second halo.}
\label{fig: Numerical effects}
\end{figure*}

In the left panel of Fig.~\ref{fig: Numerical effects} we compare the fraction of backsplashers with respect to field haloes in ESMDPL and VSMDPL over the last 6 Gyr, which corresponds to the interval over which we were able to analyze the VSMDPL simulation. Consequently, it is unknown how many passages they performed before this period, so here we use haloes that passed through another halo at least once. This is consistent with the definition used in \cite{Diemer_2021}, which we compare with our results below. We find that each fraction has a peak corresponding to haloes with 100 particles. If the peak is purely the result of numerical effects, it should depend only on the number of particles, and not their mass. In Fig.~\ref{fig: Numerical effects}, we see that the maxima of both curves correspond approximately to haloes with 100 particles, regardless of the particle mass in the simulations. The location of this peak is likely to be the result of numerical errors in tracking low-mass haloes, which has been the subject of considerable scrutiny in recent years \cite{van_den_Bosch2017, van_den_Bosch:Ogiya2018, Newton:Cautun:Jenkins:2018}. Therefore, we conclude that one can trust the results from the ESMDPL simulation for Hermeian and backsplash haloes that consist of more than 100 simulation particles at z=0. The use of less massive haloes does not change the results qualitatively but they should be used with caution.

Comparing the fractions of backsplash haloes as a function of halo mass in the two simulations, they have a common shape but differ systematically by approximately 2 per cent in absolute value, with fraction in VSMDPL being slightly higher than in ESMDPL (see the left panel of Fig.~\ref{fig: Numerical effects}, note that for a one-to one comparison in terms of fraction of backsplash haloes, one should shift the horizontal axis by the particle mass ratio of the simulations). This may be caused by the constraints placed on the large-scale distribution of matter in ESMDPL. As we show in Section \ref{sec:env}, the fractions of backsplash and Hermeian haloes depend on the density of the environment, so differences in the large-scale structure may affect these quantities. The simulations also have different snapshot frequencies, which will affect the number of backsplash and Hermeian haloes that are identified because, as discussed earlier, backsplash and Hermeian haloes that pass through targets between snapshots will not be identified by the algorithm. %This result may be explained by the fact that the ESMDPL has a too small simulation box: this leads to a distortion of the halo mass function, and also affects the frequency of halo mergers \cite{Bagla:Ray:2005}.
%This result may be explained by the fact that the VSMDPL simulation has smaller time intervals between snapshots than ESMDPL, which reduce the number of undetected passages. Another possible explanation for this discrepancy is that ESMDPL is a constrained simulation, which affects the large scale distribution of matter. As we show in Section \ref{sec:env}, the fraction of backsplash and Harmeian haloes depend on the density of environment, so differences in the large scale structure may impact the fraction of backsplashers.

Another source of difference between the backsplash fractions in ESMDPL and VSMDPL is the difference in the time between snapshots ($\lesssim 0.7$ Gyr for ESMDPL and $\lesssim 0.4$ Gyr for VSMDPL). One would anticipate that there may be haloes that enter and exit a second halo during the time between successive time steps (in this case $\texttt{pid} = -1$ and $\texttt{desc\_pid} =-1$). Moreover, one might also expect that some subhaloes leave their first host and enter another halo during this time, such that they are subhaloes in the first and second timesteps ($\texttt{pid} \neq -1$ and $\texttt{desc\_pid}~\neq -1$). The time intervals between snapshots in both simulations are smaller than the crossing time of Milky Way-sized haloes (see \cite{Sung-Ho}, for details about the estimation), so the amount of missed passages is low, yet, it could be responsible for the observed 2\% difference between the fractions of backsplash haloes.
%Despite these cases (as well as more sophisticated scenarios that our algorithm does not take into account), we are confident that this algorithm misses very few haloes that pass through other haloes. First, the simulation time steps ($\lesssim 0.7$ Gyr for ESMDPL and $\lesssim 0.4$ Gyr for VSMDPL) are smaller than the crossing time of Milky Way-sized haloes (see \cite{Sung-Ho}, for details about the estimation). %Secondly, we checked the algorithm by comparing fractions of haloes that have passed through another halo at least once in two simulations with different time steps (see Section~\ref{section: simulations with different resolution}). We also compared our results with the previous work: no significant difference was observed in both cases (namely, we see the same trend, Fig.~\ref{fig: Numerical effects}). 

\subsubsection{Comparison with the fraction of backsplash haloes from the literature}
As was pointed out in Section~\ref{section: Definitions of halo classes}, several definitions of backsplash haloes have been used by different authors. This may cause difficulties when comparing different studies. Thus, we compare the fractions of backsplash haloes that were found in the ESMDPL simulation with the results of a previous study \cite{Diemer_2021}, in which the backsplash haloes were defined in a similar manner to the one used in this paper. In order to be consistent with \cite{Diemer_2021}, here we use haloes that passed through another halo at least once, which differs from our definition given in the Section~\ref{section: Definitions of halo classes}.

Our results are compared with the results of \cite{Diemer_2021} on the right panel in Fig.~\ref{fig: Numerical effects}, where the fractions of backsplash haloes are shown with respect to all field haloes as a function of mass. At masses greater than $3 \times 10^{10} h^{-1}M_{\odot}$ our results are consistent with those of  \cite{Diemer_2021} (in the mass bins where the error bars are smaller than the difference of backsplash fractions, the relative difference is not greater than 8 per cent, and the absolute difference is less than 1 per cent). This discrepancy may be due to the chosen normalization: we have converted the Diemer bound-only mass, $M_{\mathrm{200m, bnd}}$, to mass used in our paper by assuming a NFW profile with concentration, $c_{326} = 10$. 

The sharp decline in the fraction of the backsplash haloes in ESMDPL below mass $M \sim 10^{7.5}h^{-1}M_{\odot}$ cannot be explained by any physical effect because there are no characteristic mass scales in the hierarchical model of structure formation. Over the mass range covered, the number of dark matter haloes is a power-law function of mass. Therefore, the mass dependence of the backsplash halo fraction should be monotonic. As we discussed previously, this drop is due to numerical resolution. To summarise, the right panel of Fig.~\ref{fig: Numerical effects} not only demonstrates that our findings and the results of \cite{Diemer_2021} follow the same trend, it also illustrates that ROCKSTAR tends to lose track of low-mass backsplash haloes. Many backsplash haloes with $M_{\mathrm{vir}} \lesssim 3 \times 10^{7} \;h^{-1}\;\mathrm{M_{\odot}}$, i.e. with less than $\sim 100$ particles within the virial radius, are not identified correctly.

\hfill \break
Taking into account the results discussed in Section~\ref{section: Numerical effects}, in Section~\ref{section:  Hermeian halo population} we examine only Hermeian haloes that contain more than 24 particles at the moment of the first passage through another halo and have more than 100 particles within their virial radius at $z = 0$.
%\begin{figure*}
%\centering\includegraphics[width=\textwidth]{Numerical effects.pdf}
%\caption{\textit{Left panel:} Comparison of the fractions of backsplash haloes obtained with different simulations. The backsplash haloes have been used that have performed a passage only within the last 6 Gyr. Each point corresponds to the number of backsplash haloes normalised  by the total number of field haloes more massive than the given mass. \textit{Right panel:} Comparison of the backsplash halo fraction in the ESMDPL dependently on the halo mass with the results of the \cite{Diemer_2021} work. It is important to note that, unlike the left panel, where we count all passages of haloes more massive than the given mass (i.e. the cumulative function), here the fraction of the backsplash halo in each mass bin is shown. For convenience, in both panels there are scales showing the number of particles for used simulations.}
%\label{fig: Numerical effects}
%\end{figure*}

\subsection{Hermeian halo population}
\label{section: Hermeian halo population}
In this section we report on the general properties of the Hermeian population that was found in the ESMDPL simulation, their targets, and the large-scale environment in which Hermeians occur. Note that all comparisons are carried out at z = 0 only.
\subsubsection{Hermeian halo fraction and mass function}

\begin{figure*}
\centering\includegraphics[width=\textwidth]{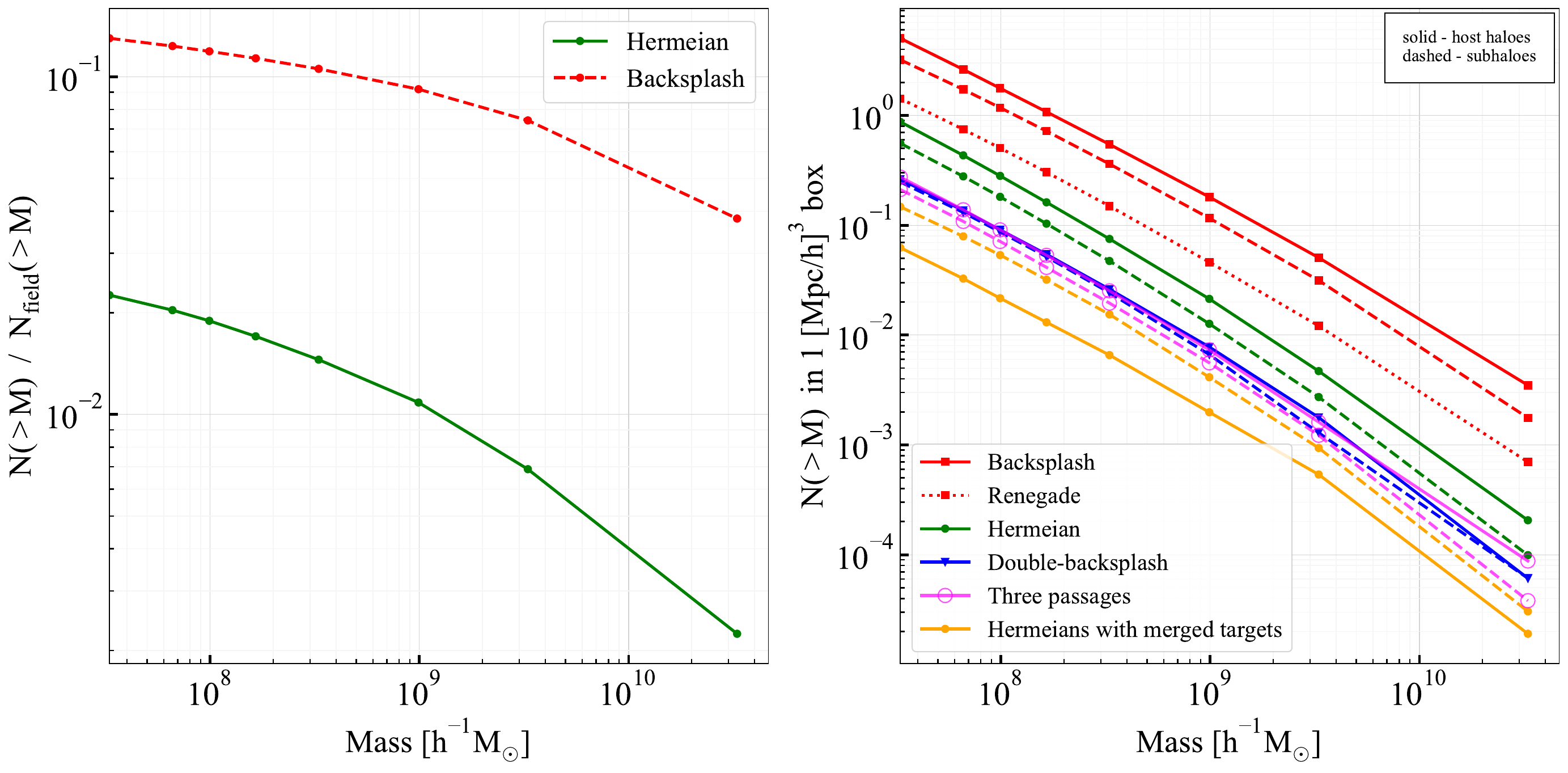}
\caption{\textit{Left panel:} Cumulative fractions of backsplash and Hermeian haloes as a functions of mass in the ESMDPL simulation at z = 0. 
\textit{Right panel:} Cumulative number of haloes of various classes with interactions in the past. Solid lines represent field haloes: backsplashers, Hermeians, double backsplashers, haloes with three passages and Hermeian-like haloes with two targets which have merged at $z=0$. Dotted line represents renegade haloes, dashed lines represent analogues of backsplashers, Hermeians, double backsplashes, Hermeian-like haloes with merged targets and haloes with three passages, but these haloes are subhaloes at $z=0$.
}
%\color{red}Cumulative number of haloes of various classes.  "Backsplash N=2" are haloes that have passed through the same target twice (i.e. the \texttt{ID} of their first targets are equal to the \texttt{ID} of their second targets at the moment of the second passage). If a backsplasher (a halo that passed through another halo only once) becomes a subhalo of a halo, which differs from the descendants of the target, we denote this as ``backsplash N=1 two targets (s)'', but this can also be called a renegade halo, since it resembles a population found in \cite{Knebe11b}.``Hermeians with merged targets'' are haloes that have passed through the two different (at the moment of the second passage) targets, however, their targets merged together by z = 0. In some way, they are "unlucky" Hermeian haloes. In this particular plot "Hermeian haloes" are haloes that have passed through the two different targets and the targets are still different at z = 0. "Backsplash haloes" are haloes that have passed through another halo only once. \textbf{"Three passages" denotes haloes that have performed three passages (we know nothing about their targets).}  The meanings of the "f" and "s" symbols are the same for all halo classes: "f" corresponds to field haloes and "s" corresponds to subhaloes at z = 0.}\color{black}

\label{fig: Fractions of haloes}
\end{figure*}

We compute the cumulative fraction of field haloes classified as backsplash or Hermeian by dividing  the number of backsplash or Hermeian haloes, $N(>M)$, by the cumulative number of all field haloes as a function of mass, $N_\mathrm{field}(>M)$, (see Fig.~\ref{fig: Fractions of haloes}, left panel). The fraction of field haloes classified as Hermeian decreases slowly as a function of mass from almost 2.3 per cent for haloes with $M_{\mathrm{vir}} > 3.3 \times 10^{7}\; h^{-1}\;\mathrm{M_{\odot}}$ to 0.4 per cent at $10^{10}\; h^{-1}\;\mathrm{M_{\odot}}$. At low masses, Hermeian haloes are approximately 6 times less numerous than backsplash haloes. The size of the difference increases with halo mass.

Besides backsplash and Hermeian haloes there are many other evolutionary pathways of haloes that have experienced interactions with other haloes. We compare their abundances in the Universe on the right panel of Figure~\ref{fig: Fractions of haloes}. The most abundant in this `zoo' are backsplash haloes, which are followed by haloes that passed through a target and then became a subhalo, maybe in the same target. If a backsplasher becomes a subhalo of a halo, which differs from the descendants of the target, we denote this as ``Renegade'', since it resembles a population found in \cite{Knebe11b}. We also present in this Figure the mass function of backsplashers that interacted with the same halo twice, as these objects resemble Hermeians by the number of interactions.
The haloes that experienced three passages in the past through 1, 2 or 3 targets are less abundant than Hermeian haloes, which is expected: the probability to encounter more targets decreases with the number of targets. Finally, there is a small population of haloes that passed through two distinct targets that are not distinct by $z=0$.
Among the field haloes in Fig.~\ref{fig: Fractions of haloes} (solid lines), Hermeian population is the second largest after the backsplahsers.

Several of the halo classes in Fig.~\ref{fig: Fractions of haloes} have interacted closely with more than one target: renegade, Hermeians, double backsplashers that end up as subhaloes, and haloes with three pasages. All of these types of halo could facilitate the exchange of material between host haloes. This fact makes the Hermeian and other similar halo populations important participants in the evolution of their targets. The transfer of material between targets by Hermeian haloes has been demonstrated in \cite{Newton}. Further study of matter transfer in hydrodynamical simulations is needed to quantify the role of this process in the evolution of galaxies. Our results in Fig.~\ref{fig: Fractions of haloes} show that renegade and Hermeian populations should be the considered for these studies, as they are the most numerous.

\subsubsection{Abundance of targets and target masses}

\begin{figure*}
\centering\includegraphics[width=\textwidth]{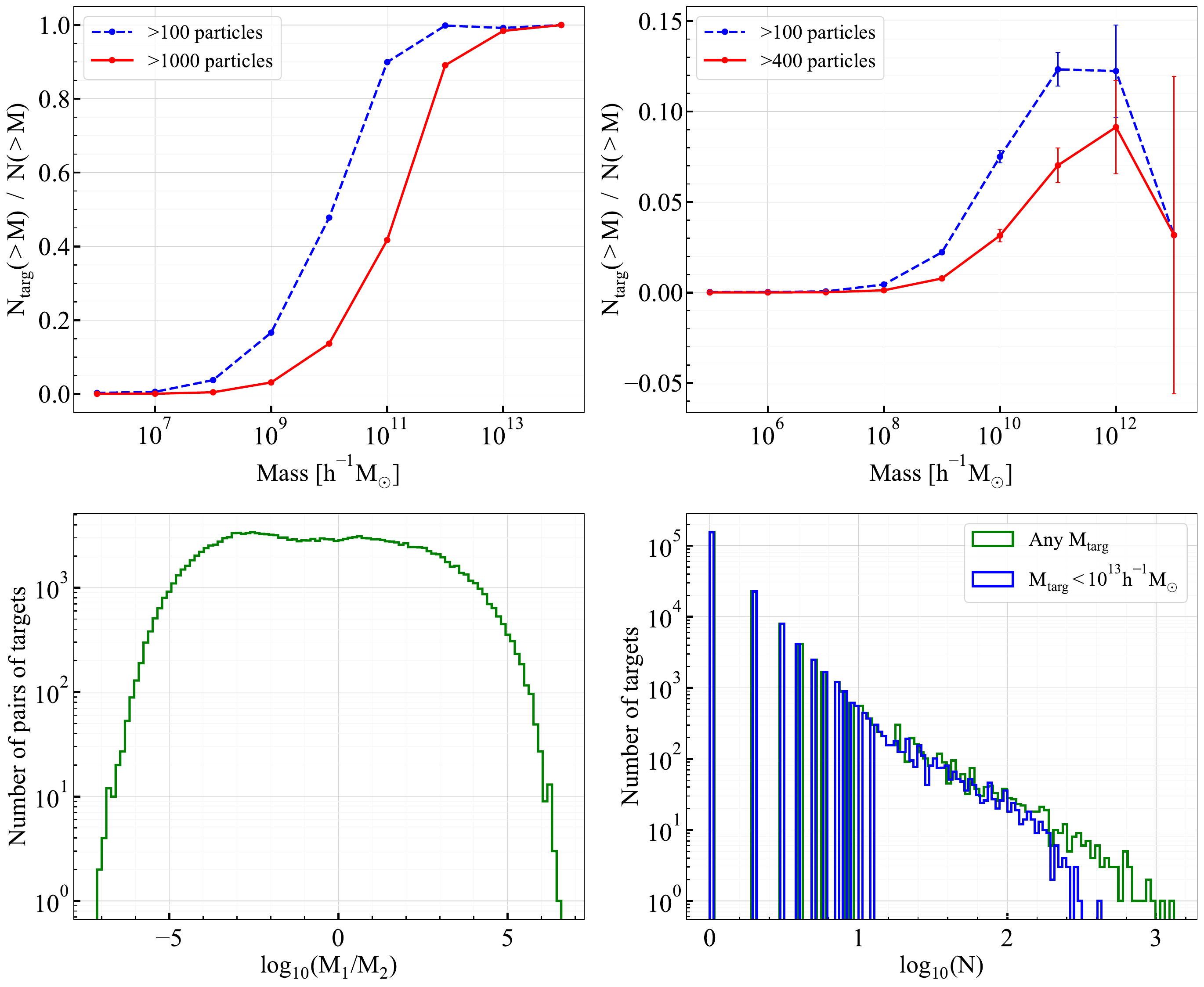}
\caption{Characteristics of systems with Hermeian haloes found in the ESMDPL simulation. \textit{Top left:} Cumulative fraction of target haloes (i.e. descendants of haloes that have been passed through by Hermeians) as a function of halo mass for the Hermeian haloes consisting of more than 100 simulation particles at z = 0 (dashed line) and for the Hermeian haloes consisting of more than 1000 simulation particles at z = 0 (solid line). \textit{Top right:} Fraction of targets under mass ratio restriction $0.5<M_1/M_2<2$ at z = 0.\textit{Bottom left:} Histogram of the z = 0  mass ratio of the pairs of target halo. \textit{Bottom right:} Histogram of the number of passages through a target for all targets as well as for the targets with masses limited to $M<10^{13}\; h^{-1}\;\mathrm{M_{\odot}}$.}
\label{fig: Targets}
\end{figure*}

In order to study if Hermeian haloes can participate in the exchange of matter in a LG-like configuration, we examine what fraction of haloes have participated in a Hermeian encounter as a target.
The upper-left panel of Figure~\ref{fig: Targets} shows that more than half of haloes with $M_{\mathrm{vir}} > 10^{11}h^{-1}M_{\odot}$  have participated as targets for Hermeian haloes. %from a certain mass, the first larger part and then all the haloes in the simulation present as targets.
Of note, more than 95 per cent of all haloes with $M_{\mathrm{vir}} > 10^{12}\; h^{-1}\;\mathrm{M_{\odot}}$ have been targets at least once (i.e. a halo passed through them at least once). This suggests that nearly all LG-like objects have experienced Hermeian encounters. 
The fraction of targets at the high-mass end saturates above certain mass, which shifts to smaller masses when the mass threshold of Hermeian haloes decreases.
%does not depend on the minimal mass threshold of Hermeian haloes which passed through these targets. 
To demonstrate this, we increase the mass threshold by 10 times, i.e. from $3.3 \times 10^{7}\; h^{-1}\;\mathrm{M_{\odot}}$ to $3.3 \times 10^{8}\; h^{-1}\;\mathrm{M_{\odot}}$. These results suggest that in simulations with sufficiently high resolution, the haloes of any mass will be targets. In other words, the Hermeian haloes may be present around any halo.
%\subsubsection{Mass ratio between targets}

The lower-left panel of Figure~\ref{fig: Targets} presents the distribution of the mass ratio of the pairs of haloes which make Hermeians. $M_{1}$ and $M_{2}$ refer to the z = 0 masses of the first and second halo the Hermeian encountered, respectively. The histogram is asymmetric\footnote{We investigated whether this was affected by the choice of bin size but found no qualitative difference.}: on average, the first target tends to be slightly less massive than the second.% The number of cases with the target mass ratio (without a direction) ranging within $1 - 1000$ are approximately equally frequent. Thus, Hermeian haloes appear with approximately equal probability in pairs of systems, where the mass ratio is smaller than 2 whose masses differ by tens and even hundreds of times.% 
The target mass ratio distribution is wide and flat: it varies almost uniformly between $10^{-4}$ and $10^3$. The pairs with the mass ratio between 0.5 and 2 have a particular interest, since they resemble the LG, and such pairs occupy only a small fraction of this distribution.

%\subsubsection{Fraction of targets under mass ratio restriction}

To see whether target haloes with similar masses are well-connected by Hermeian haloes, we study those systems with a target mass ratio of $0.5 < M_{1}/M_{2} < 2$. In the upper-right panel of figure~\ref{fig: Targets} we show the cumulative fraction of targets with respect to all field haloes as a function of halo mass. In contrast with the upper-left panel of Figure~\ref{fig: Targets}, only 12 per cent of all haloes more massive than $10^{12}\; h^{-1}\;\mathrm{M_{\odot}}$ have been targets at least once.

\subsubsection{Number of passages}
In the lower-right panel of Fig.~\ref{fig: Targets} we show the number of times a target has been encountered by a Hermeian halo. Of 202 thousand targets, 77 per cent (about 155 thousand) have encountered only one Hermeian halo. In the other extreme case the maximum number of Hermeians that passed through a single halo is well over 1000. The number of passages depends both on the minimum mass of Hermeians in the simulation (number of passages increase with the decrease of mass) and the mass range of the targets under consideration. 

In the study carried out by \cite{Newton}, they found that in one of their LG realizations, 121 Hermeian haloes had passed through both main LG members. In our study, the passage of more than 100 haloes through targets with masses within $[5.5\times10^{11} - 2\times10^{12}]h^{-1}\mathrm{M_{\odot}}$ are observed (35 of the 1615 targets in this mass range), but our DM particle mass and, hence, minimum Hermeian mass is two times higher than in \cite{Newton}. The passage of thousands of Hermeian haloes are associated only with the most massive haloes in the simulation with masses above $10^{14}$ $h^{-1}\,$M$_\odot$.

\subsubsection{Distance between targets and time between passages}
\begin{figure}
\centering\includegraphics[width=\linewidth]{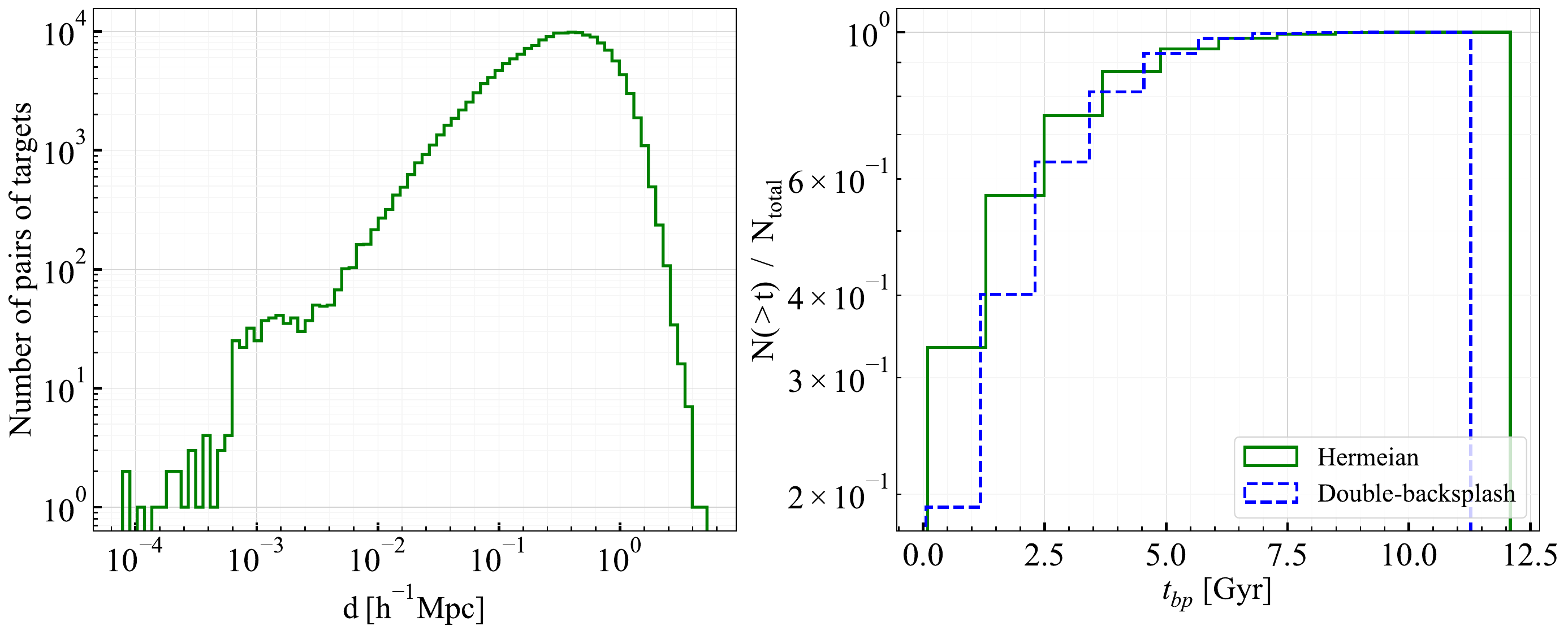}
\caption{\textit{Left panel:} Histogram of distances between Hermeian targets at z = 0. \textit{Right panel:} Distribution of the time between two passages of Hermeian and double-backsplash haloes, $t_{\mathrm{bp}}$. For Hermeians, it is the time they spend flying away from their first target before being pulled into their second (solid line). The time between two passages of double-backsplash haloes (dashed line) is defined in similar way.} %This time is defined as $t_{\mathrm{bp}}$ = $t_{\mathrm{in}} - t_{\mathrm{out}}$, where $t_{\mathrm{out}}$ is the average between the last snapshot where a Hermeian halo is still inside the first target (\texttt{pid} $\neq$ -1) and the first snapshot where it is a field halo (\texttt{pid} = -1), $t_{\mathrm{in}}$ is the average between the last snapshot where a Hermeian halo is still outside the second target (\texttt{pid} = -1) and the first snapshot where it is a subhalo (\texttt{pid} $\neq$ -1). The time between two passages of double-backsplash haloes (dashed line) is defined in similar way.}
\label{fig:Time between passages all HH}
\end{figure}

The distribution of distances between targets is shown in the left panel of Fig.~\ref{fig:Time between passages all HH}. The majority of the targets (62 per cent) are not more than 0.4 $h^{-1} \;\mathrm{Mpc}$ away from each other at z = 0. 94 per cent of targets are at a distance not greater than 1 $h^{-1} \;\mathrm{Mpc}$ from each other at redshift zero.  
In the study by \cite{Newton} approximately 90 per cent of the Hermeian haloes were found in the simulation, where the distance between the MW and M31 analogues was nearly 0.7 Mpc (in two other simulations, there were larger separations of the primary pairs). Such results are consistent with the typical separation of targets that we find in this work. %\color{red} does this depend on pair mass ratio? or total pair mass \color{black}

The time between the interactions of the Hermeian halo with the first and second targets is shown in the right panel of Fig.~\ref{fig:Time between passages all HH}. The beginning of the period between passages is calculated as the average of the time of the last snapshot when the halo is inside the first target, and the time of the first snapshot when it is outside the first target. The end of the period is the average of the times of the last snapshot before the halo enters the second target, and the first snapshot when it is inside the second target. Half of the Hermeian haloes spend more than 2.1~Gyr between the first and second targets (the average time is 2.6~Gyr). This time may be sufficient to re-accrete the gas after it has been lost during the first passage. The process of re-accretion could be studied in Hermeian haloes with large inter-target time intervals in hydrodynamical simulations. Interestingly, double-backsplash haloes spend even more time between passages through the target with median time of 2.7~Gyr (the average time is 2.9~Gyr).

%associated a large number of the Hermeians in their Local Group simulations with a group infall event.
%This study has revealed that %only one Hermeian halo passed through 75 per cent of the targets 
%75 per cent of targets have participated in a single Hermeian encounter (see the upper-right panel of Fig.~\ref{fig: Targets}). However, there are haloes that have been passed through several thousand times. Thus, one can say that a group passage described in the article by \cite{Newton} is much rarer than a single halo passage, but it can happen (\cite{Newton} found that in the 17\_11 HESTIA simulation most of the Hermeian haloes were created as part of a large group that falls into the M31 analogue at early times). 
\subsubsection{Velocities of Hermeians and backsplashers}
\label{sec:vel}
Both Hermeian and backsplash haloes have one common property: they are, in general, receding from their last target. Backsplash haloes and galaxies studied in the literature also have high velocities with respect to the surrounding field haloes \cite{Benavides21}. Hermeian haloes could have even higher velocities, since they can be accelerated by the gravitational assist from the two target haloes. We check this using the same approach as in \cite{Benavides21}: we select field haloes within 1~$h^{-1}\; \mathrm{Mpc}$ around every Hermeian, backsplash, and double-backsplash halo and calculate the average absolute value of the peculiar velocity difference between the neighbour and the Hermeian, backsplash or double-backsplash halo. For backsplash haloes we find the median value of 175~$\mathrm{km\;s^{-1}}$ (with the sample mean and standard deviation of 214~$\mathrm{km\;s^{-1}}$ and 152~$\mathrm{km\;s^{-1}}$, respectively), close to the value of 200~$\mathrm{km\;s^{-1}}$ in \cite{Benavides21}. For double-backsplash haloes we find the median value of 202~$\mathrm{km\;s^{-1}}$ (with the sample mean and standard deviation of 235~$\mathrm{km\;s^{-1}}$ and 148~$\mathrm{km\;s^{-1}}$, respectively). For Hermeian haloes, the median velocity with respect to the environment is 239~$\mathrm{km\;s^{-1}}$ (with the sample mean and standard deviation of 269~$\mathrm{km\;s^{-1}}$ and 159~$\mathrm{km\;s^{-1}}$, respectively), which is 36 per cent higher than for backsplash haloes. 6 per cent of backsplashers, 6.4 per cent of double-backsplashers, and 9.5 per cent of Hermeians have velocities higher than 500~$\mathrm{km\;s^{-1}}$ with respect to the environment. This demonstrates that Hermeian haloes are somewhat faster than backsplash haloes.

Their higher velocities allow Hermeian haloes to travel further. The median distance between a Hermeian halo and its last target is 0.34~$h^{-1}\; \mathrm{Mpc}$, while for backsplashers and double-backsplashers this distance is 0.24 $h^{-1}\; \mathrm{Mpc}$ and 0.23~$h^{-1}\; \mathrm{Mpc}$, respectively\footnote{The sample mean $\pm$ standard deviation for these samples are: $0.44\pm 0.36$~$h^{-1}\; \mathrm{Mpc}$ for Hermeians, $0.34\pm 0.33$~$h^{-1}\; \mathrm{Mpc}$ for backsplash haloes, and $0.31\pm 0.28$~$h^{-1}\; \mathrm{Mpc}$ for double-backsplashers.}. According to \cite{Diemer_2021}, about 60\% of backsplash haloes with $M\sim10^{10}$~M$_\odot h^{-1}$ are indeed subhaloes on elongated orbits and they are gravitationally bound to their targets. We expect the fraction of bound haloes is even higher for double-backsplashers.

\subsubsection{Concentration}

The internal structure of dark matter haloes is a question of great interest because galaxy formation and evolution are based on it, it plays an important role in searches for dark matter annihilation products, etc \cite{Newton, SánchezConde:Prada:2014, Evans04}. For the analysis of halo concentration, we assume an NFW density profile \cite{NFW1996}. It is important to mention that not all haloes are perfectly described by the NFW profile, and in some cases the Einasto profile provides a more accurate approximation \cite{Einasto1965,Navarro_Hayashi_2004, Gao:Liang:2008, Navarro:Ludow:2010, Dutton:Maccio:2014}. However, for low-mass haloes, these approximations are very similar within the virial radius  \cite{Klypin:Yeps:Gottlober:2016}. To analyze the concentrations of the Hermeians, it is more convenient to use the NFW profile since, for a given halo mass, the concentration parameter uniquely determines the distribution of the dark matter inside a halo. Adopting the Einasto profile introduces an extra parameter that affects the halo density profile and, consequently, their concentration. %In addition, the use of a more accurate profile does not lead to better results: in practice, the Einasto approximation causes significant fluctuations in the resulting concentration \cite{Prada:Klypin:2012}.

\begin{figure}
\centering\includegraphics[width=\linewidth]{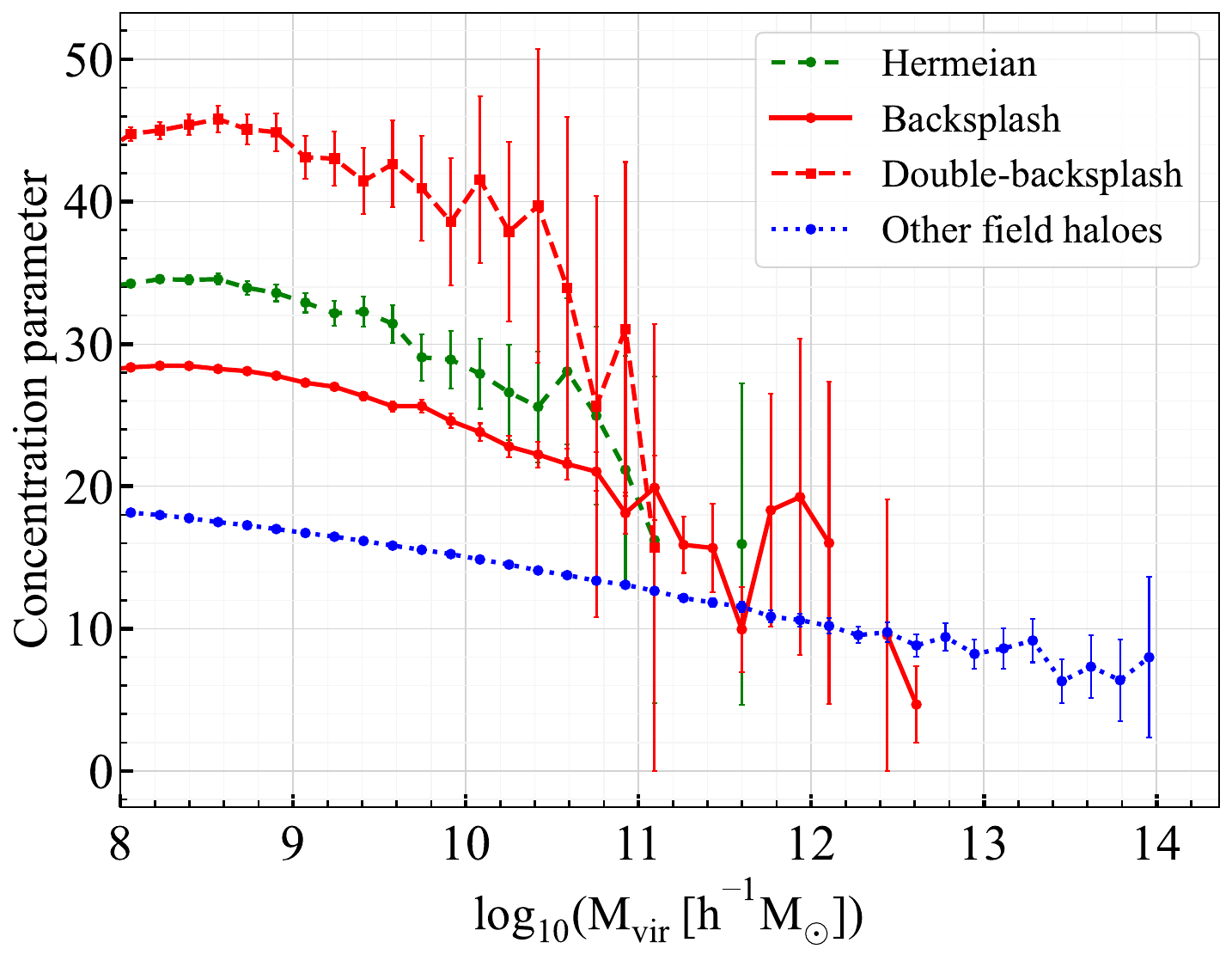}
\caption{Median concentration values obtained through the profile-independent method for Hermeians, backsplashers, double-backspleshers and other field haloes dependent on halo mass at z = 0.}
\label{fig:Concentration}
\end{figure}

There are several methods that can be used to determine the concentration of a dark matter halo. One of the standard approaches is to fit an analytical profile, followed by the determination of the concentration parameter as the ratio of the virial radius to the characteristic scale radius of the fitted profile. However, this method can be quite complicated \cite{Prada:Klypin:2012}. The comparison of halo concentrations estimated by direct fitting the NFW approximation to halo density profiles with concentrations obtained using a profile-independent method, gives similar results for all haloes, except the most massive ones \cite{ Dutton:Maccio:2014, Prada:Klypin:2012}. The profile-independent method here means finding the concentrations through the ratio of the virial velocity to the maximum of the circular velocity: 
\begin{equation}
    \left( \frac{V_{max}}{V_{vir}} \right)^2 = \frac{0.216\,c}{\ln (1+c)- c/(1+c)}.
\end{equation}
This approach does not require the fitting procedure of the halo density profile by a certain analytical one, and in the general case, the assumption of the profile type is also not needed \cite{ Prada:Klypin:2012, Klypin:Trujillo-Gomez:2011, Pilipenko:Sanchez-Conde:2017}. 
Thus, we use concentrations defined through the ratio of the maximal and virial circular velocities, found by ROCKSTAR. These concentrations may be unreliable if the halo profile is truncated by the stripping of the material at radius $\mathrm{R<R_{vir}}$, since this truncation affects only the virial velocity and increases the concentration, while the central part of the halo does not change at all. However, both Hermeian and backsplash haloes in our simulations have spent many dynamical times outside the target haloes where the stripping takes place, so they have enough time to re-virialise, which restores the NFW profile even at the periphery of the haloes. This has been explicitly tested in \cite{Newton}, so the concentrations measured in this work reflect the changes to the physical density profiles and are not just artifacts arising from the truncation.

The concentration characterizes the density at the halo center, and the dark matter annihilation signal increases in approximate proportion to the concentration in the power of 2\footnote{This holds for concentrations in the range $c=4-100$ with the deviation below 40\%.} \cite{Newton}. 
Our analysis shows that over the mass interval where there are enough haloes such that the statistical errors are small, the concentrations of Hermeians are almost 2 times higher than the concentrations of field haloes, and backsplash concentrations are approximately 1.5 times higher with respect to the remaining field haloes (see Fig.~\ref{fig:Concentration}), in agreement with previous work on the backsplash and Hermeian populations in the Local Group \cite{Newton}. We find that double-backsplash haloes are nearly 2.5 times more concentrated than the remaining field halo population and 1.3 times more concentrated than the Hermeians. As we pointed out in Subsection \ref{sec:vel}, we expect that most of the double-backsplashers are indeed subhaloes on elongated orbits. Subhaloes have higher concentrations than field haloes, which explains why they have higher concentrations than the Hermeians.

\subsubsection{Large-scale environment}
\label{sec:env}
The large-scale environment has been shown to affect galaxy interactions such as mergers and flybys \cite{An:Kim:Moon:2019, L'Huiller:Park:Kim:2015}. There is evidence to suggest that the rate of multiple interactions also depends on the environment density at the considered location.  In order to study how the density of the halo environment affects the number of Hermeians and backsplashers, we compare the dependence of the large-scale overdensity on halo mass for Hermeian, backsplash, and the remaining field halo populations at z = 0. Various methods are frequently employed to define environment \cite{Muldrew:Croton:Skibba:2011}. In this paper, we define the large-scale overdensity using a grid-based Cloud-In-Cell (CIC) smoothing with a voxel size of 64 $[h^{-1}\; \mathrm{Mpc}]/128$ = 0.5 $h^{-1} \;\mathrm{Mpc}$. However, a previous study demonstrated that a density field with a characteristic resolution of 8 times the halo virial radius provides a reliable picture of the environment \cite{ Lee:Primack:Behroozi:2016}. Therefore, with our basic resolution, 0.5 $h^{-1}\; \mathrm{Mpc}$,  this condition is satisfied for all haloes with a mass $\lesssim 5 \times 10^{10} \; h^{-1}\; \mathrm{M_{\odot}}$, which is large enough for most Hermeians and backsplashers. Nevertheless, to consider MW-sized haloes, we need to smooth the box on scales of approximately 1.6 $h^{-1}\; \mathrm{Mpc}$.  For this reason, according to the procedure used by \cite{ Lee:Primack:Behroozi:2016}, we employ a Gaussian smoothing approach in which we convolve the CIC density cube with a 3D Gaussian kernel. The kernel has the same number of voxels as the CIC cube and has a half-width at half-maximum (HWHM) equal 1.6 $h^{-1}\; \mathrm{Mpc}$. %A major advantage of this method of determining local density is that it provides the density field at each point in the simulation volume in one cycle through all haloes, in addition, when the environment density at the halo location is required, one may find its value in the generated array just using ROCKSTAR coordinates of the analyzed halo.
\begin{figure}
\centering\includegraphics[width=\linewidth]{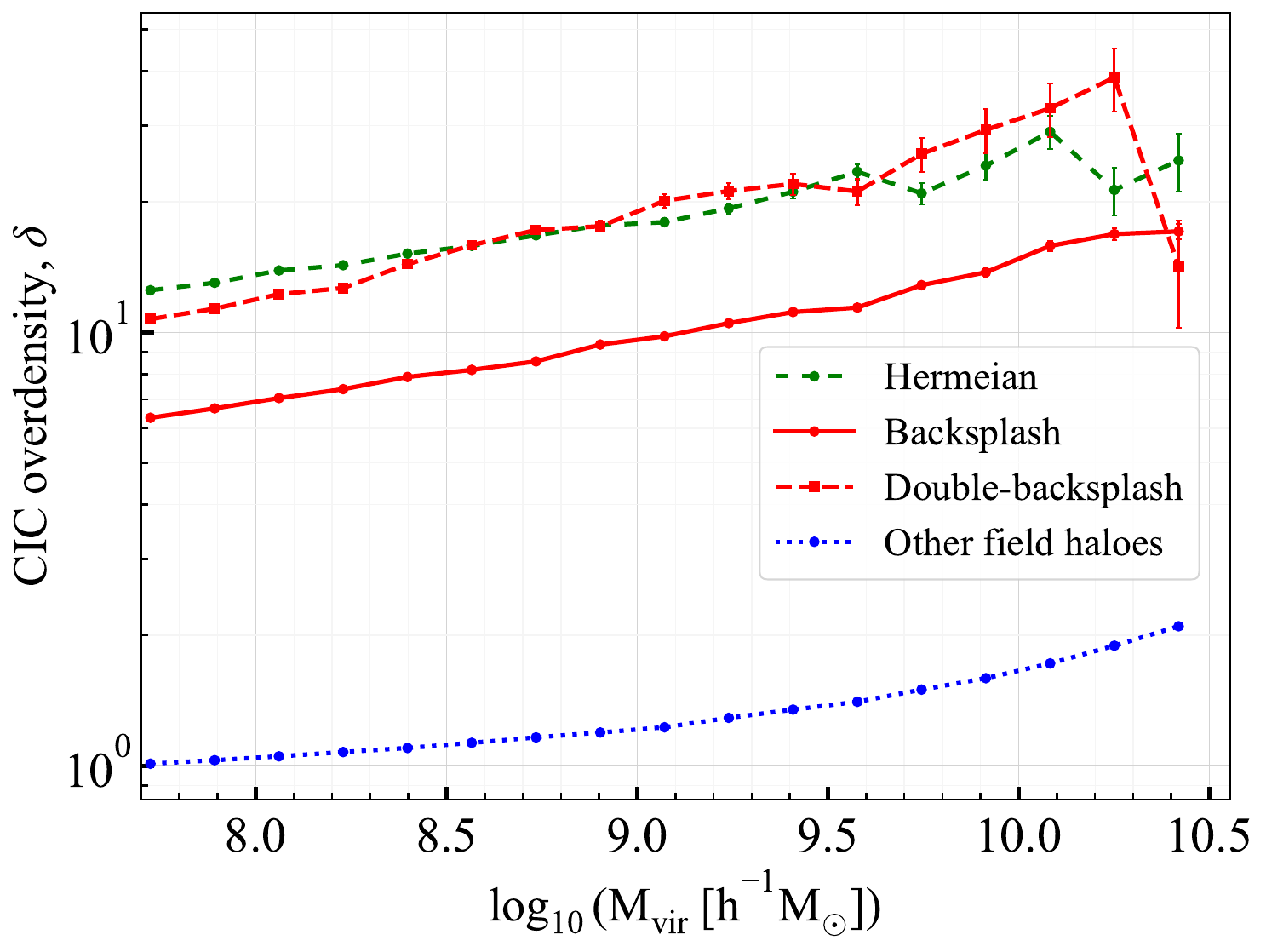}
\caption{Dependence of the median CIC (grid-based Cloud-In-Cell) overdensity on halo mass for Hermeian, backsplash, double-backsplash and remaining field haloes at z = 0. The overdensity is given in units of the mean density. To compute the environmental density we implement CIC smoothing with a voxel size of 0.5 $h^{-1}\; \mathrm{Mpc}$.}
\label{fig: Density}
\end{figure}

\begin{figure*}
\centering\includegraphics[width=\textwidth]{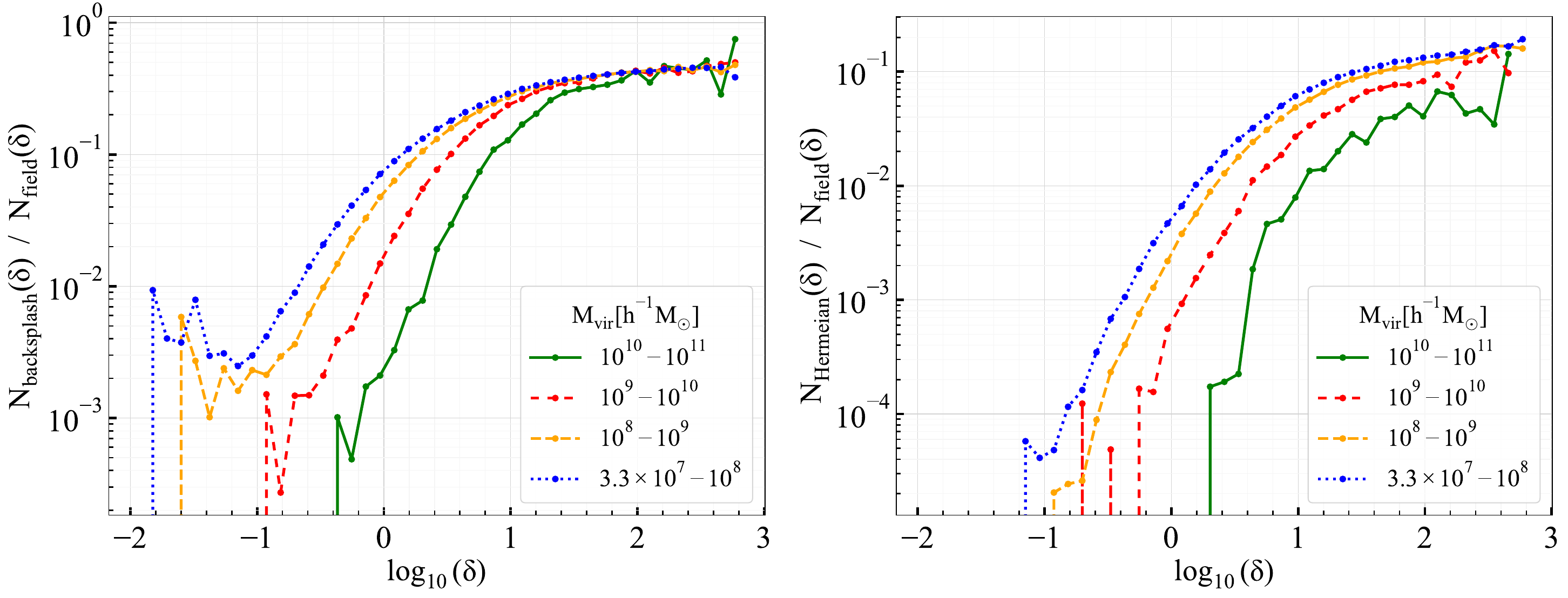}
\caption{Dependence of the fractions of  backsplash (left panel) and Hermeian (right panel) haloes on the CIC overdensity at z = 0 ($\delta$). The overdensity is given in units of the mean density. To compute the environmental density we implement CIC smoothing with a voxel size of 0.5 $h^{-1}\; \mathrm{Mpc}$. In order to highlight the dependence of the analyzed halo fractions on the environment and remove the dependence on the mass, we divide the haloes into four groups according to their mass, see the legend for the specific values (mass in units $h^{-1}\;\mathrm{M_{\odot}}$).}
\label{fig: Density_revers}
\end{figure*}
Figure~\ref{fig: Density} shows that the environmental overdensity around Hermeian and backsplash haloes is significantly higher than around other field haloes. In order to check whether more Hermeians and backsplashers tend to appear in overdense regions than in underdense ones, we study their fractions as functions of the CIC overdensity. From Fig.~\ref{fig: Density_revers}, it can be seen that regardless of the halo mass, the fractions of both Hermeian and backsplash haloes first increase with the environment overdensity and then reach a constant value. The majority of haloes passing through other haloes are found in overdense regions, despite the fact that these kinds of halo interactions occur in a wide range of density environments.
% \begin{figure*}
% \centering\includegraphics[width=\textwidth]{Hermeians_web_black.tiff}
% \caption{}
% \label{fig: Hermeian web}
% \end{figure*}

\begin{figure*}
\centering\includegraphics[width=\textwidth]{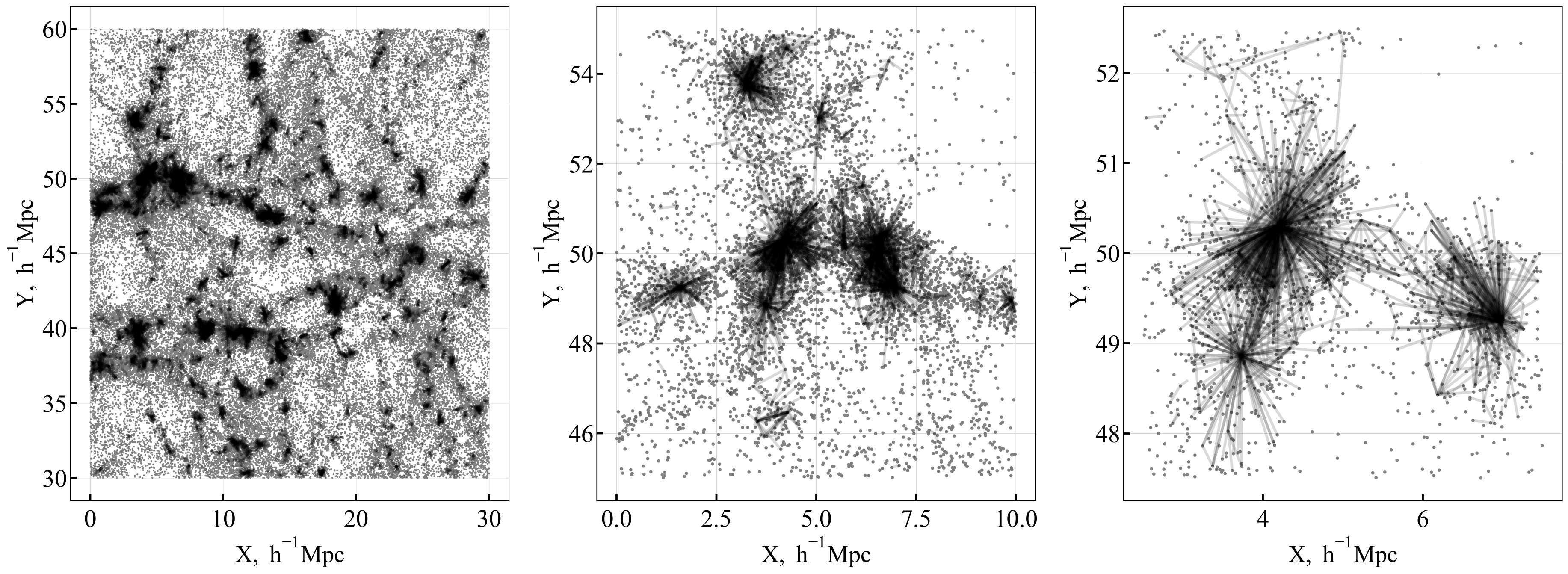}
\caption{Spatial distribution of target-haloes connected by Hermeians in three cubes with decreasing volume, from left to right (each line starts at the first target of a Hermeian halo and ends at its second target). All field haloes more massive than $10^{9}$ $h^{-1}\,$M$_\odot$ are shown.}
\label{fig: Hermeians web}
\end{figure*}

%\begin{figure*}
%\centering\includegraphics[width=\textwidth]{Hermeians_web_blue_red.pdf}
%\caption{.}
%\label{fig: Hermeians web2}
%\end{figure*}

Another way to represent the impact of the environment is to describe the properties of backsplash and Hermeian haloes in the direct vicinity of other haloes \cite{ Gill:Knebe:Gibson:2005,Moore04, Ludlow09}. The fraction of  field haloes that are Hermeian within 1-2~$\mathrm{R_{vir}}$ of all field haloes with masses greater than $5.5\times10^{11} h^{-1}\;\mathrm{M_{\odot}}$ (the lowest mass of the Milky Way that we adopt) is 9 per cent. This decreases to 7 and 4 per cent for haloes within 1-3~$\mathrm{R_{vir}}$ and 2-3~$\mathrm{R_{vir}}$, respectively . At the same time, the fraction of field haloes that are backsplash is 42 per cent within 1-2~$\mathrm{R_{vir}}$, 35 per cent within 1-3~$\mathrm{R_{vir}}$, and 23 per cent within 2-3~$\mathrm{R_{vir}}$. 

The targets of Hermeian haloes are also distributed in a non-uniform way across the Universe. To demonstrate this we plot all field haloes  with masses above $10^{9}$ $h^{-1}\,$M$_\odot$ from a small part of the ESMDP simulation volume in Fig.~\ref{fig: Hermeians web}. Lines in this Figure connect pairs of haloes connected by Hermeians (each line starts at the first target of a Hermeian halo and ends at its second target). It can be seen that the target haloes have a patchier distribution than the field haloes. On the right panel of Fig.~\ref{fig: Hermeians web} the most of the lines come from three haloes with masses $>10^{13} h^{-1}\;\mathrm{M_{\odot}}$. So massive haloes are likely to be connected with their neighbours by a large number of lines.

%The next section of this paper focuses on Hermeian haloes of the Local Group.

\subsection{Hermeian haloes of the Local Group}
As we discussed in Section~1, the properties of Hermeian haloes make them interesting targets to search for in the Local Group. %In order to identify these objects in observations, one needs to understand how to distinguish them from the other field haloes exactly within the Local Group. 
Indeed, NGC 3109 has already been proposed as a candidate backsplash or Hermeian galaxy \cite{Newton,Banik:Haslbauer:Pawlowski:2021}, and more could await identification. In order to confirm this classification, one must understand how to distinguish Hermeians from the other field haloes within the LG. 
The first study to attempt this characterization was carried out by \cite{Newton}; however, it is limited by the small sample size of LG analogues they considered. We address this limitation by selection of LG analogues from the ESMDPL simulations, from which we obtain 108 LG-like systems.
To reveal the observational features of the Hermeian haloes in the Local Group, we selected its analogues in the ESMDPL simulation and study the associated Hermeian halo population which occurs. The following sections review our main findings. Note that all comparisons are carried out using z = 0 data only and, like in Section~\ref{section:  Hermeian halo population}, we draw our results from a sample of haloes with $\mathrm{M_{vir}} > 3.3 \times 10^{7} \; h^{-1}\; \mathrm{M_{\odot}}$  (i.e. with a mass of at least 100 simulation particles).

\subsubsection{Local Group analogues}

Since ESMDPL is a constrained simulation, by construction it contains a Local Group replica, i.e. a pair of haloes with the correct LG characteristics embedded within a large-scale environment that is consistent with the observations. We are also able to select other halo pairs with similar internal properties such as the total mass, the mass ratio, the distance between the halo pair, the relative velocity, etc. However, these are not embedded in the same large-scale environment, i.e. there is no Virgo cluster at an appropriate distance. Significant variance in the selection criteria adopted by different authors to find LG analogues can affect the results \cite{Forero-Romero:Hoffman:Bustamante:2013, Fattahi:Azadeh:Navarro:2016, Carlesi:Edorado:Hoffman:2020}. To identify LG analogues we chose the criteria from \cite{Sorce:Ocvirk:Aubert:2022}, which allows us to have a large enough sample, but LG analogues\footnote{called LG look-alike in \cite{Sorce:Ocvirk:Aubert:2022}} still resemble the real LG:
%Therefore, it is necessary to limit the parameters of systems similar to the LG carefully. However, the more restrictive the selection criterion, the smaller the number of LG analogues found in the simulation. We believe that at the initial stage of studying a new class of objects, it is more important to choose a simple selection criterion and to obtain a larger sample than to take into account subtleties of determining observed parameters of the LG. To identify LG analogues (the LG look-alike in \cite{Sorce:Ocvirk:Aubert:2022}), the following criteria for choosing pairs of haloes were used \cite{Sorce:Ocvirk:Aubert:2022}:

\begin{flushleft} 
\begin{itemize}
    \item their masses are between $5.5 \times 10^{11}$ and $2 \times 10^{12}\; h^{-1}\; \mathrm{M_{\odot}}$,
    \item their separation is smaller than $1.5\; h^{-1} \;\mathrm{Mpc}$,
    \item their mass ratio is smaller than 2,
    \item there is no other halo more massive than $5.5 \times 10^{11} h^{-1}\;\mathrm{M_{\odot}}$ within a sphere of radius 1.5 $h^{-1}\; \mathrm{Mpc}$ centered on the most massive halo in the pair.
\end{itemize}
\end{flushleft} 
Within each pair, we consider the less massive halo as the Milky Way analogue, and the more massive halo as the Andromeda Galaxy analogue (M31). 

In the section that follows, we present the summary statistics for selected LG-like systems in the ESMDPL simulation.
\subsubsection{Statistics}

\begin{figure}
\centering\includegraphics[width=\linewidth]{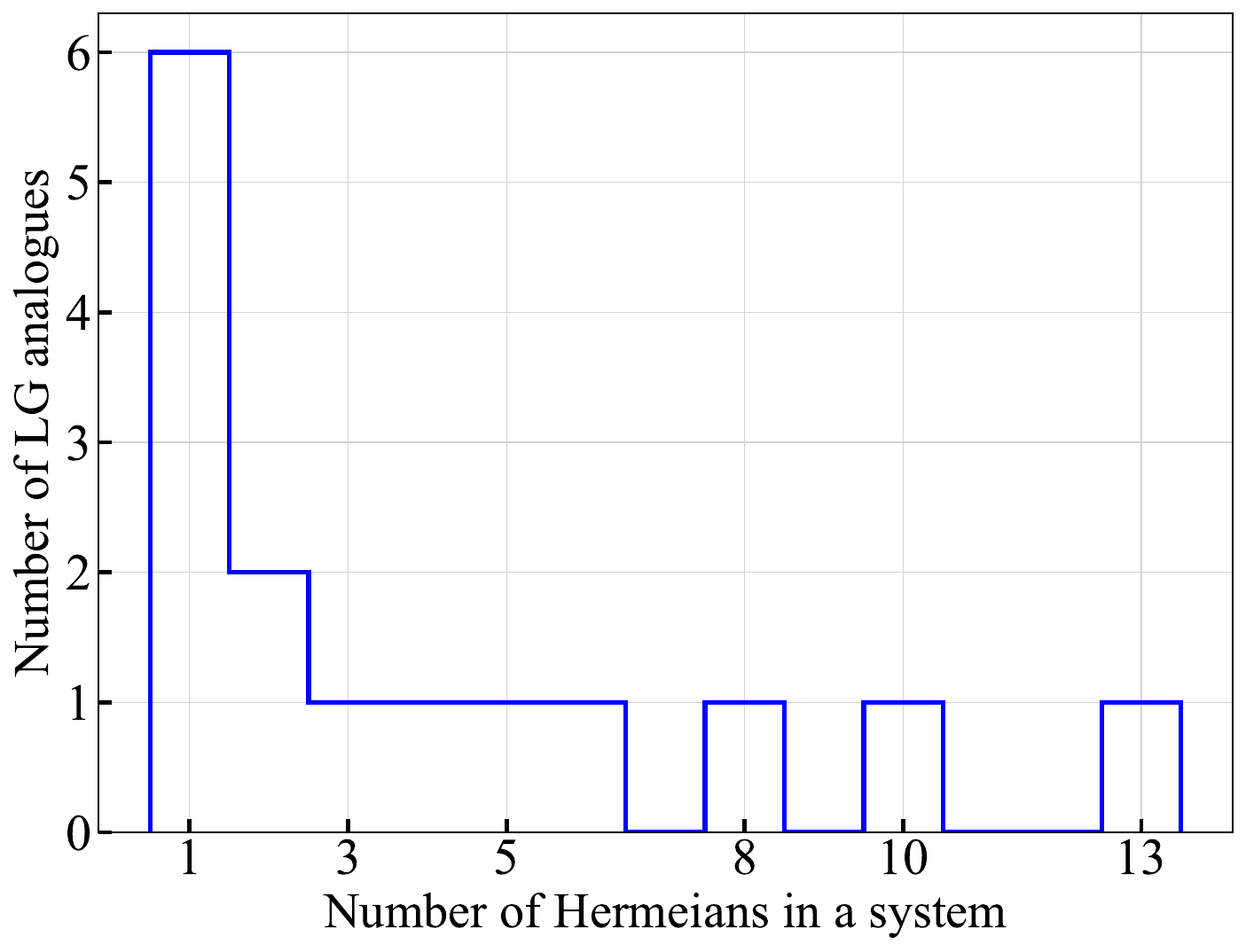}
\caption{Histogram of the number of Hermeian haloes in 15 out of 108 LG-like systems, which host Hermeian haloes with mass $\mathrm{M_{vir}} > 3.3 \times 10^{7} \;h^{-1}\; \mathrm{M}_\odot$. %\textit{Right panel:} The time between passages defined similar to Fig.~\ref{fig:Time between passages all HH}, but now only for Hermeians from LG analogues.
}
\label{fig: Number of Hermeian haloes in LG and t_bp}
\end{figure}

\begin{figure*}
\centering\includegraphics[width=\linewidth]{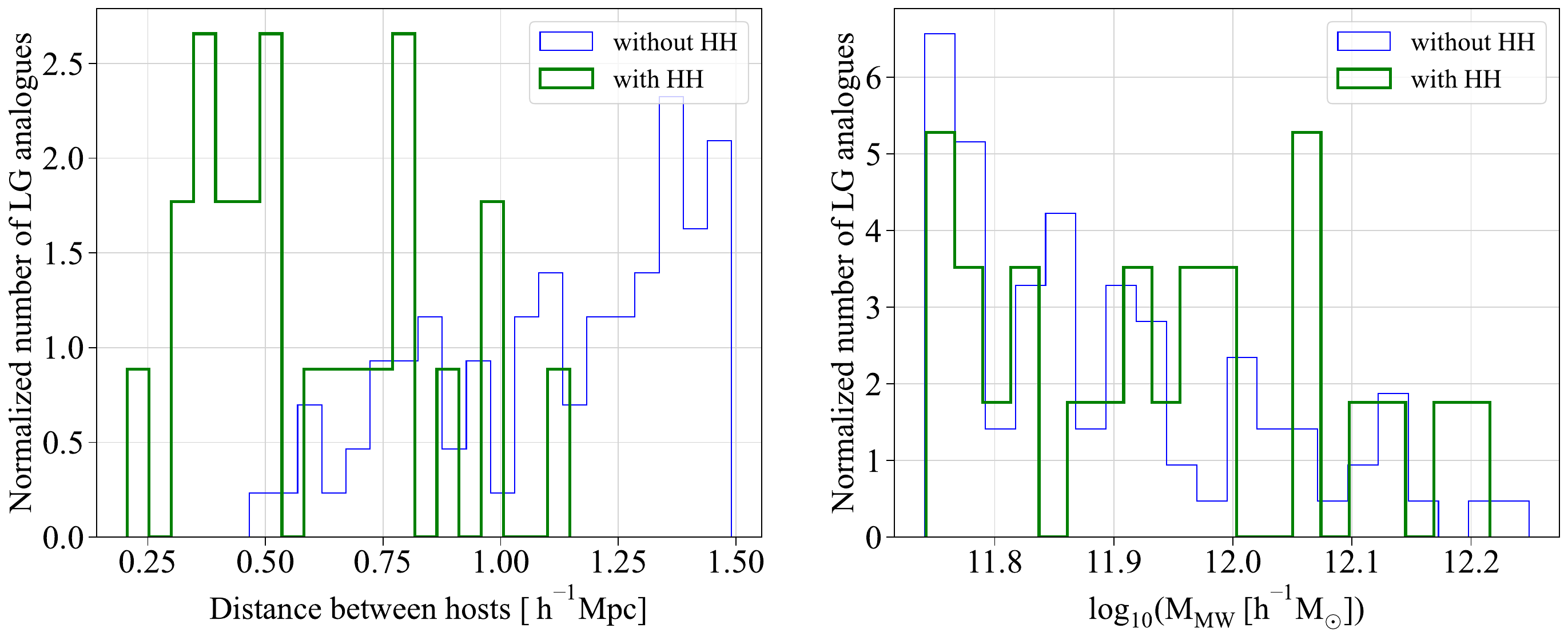}
\caption{\textit{Left panel:} Distribution of distance between hosts in LG-like systems with and without Hermeian haloes (HH) at z = 0. The chance for LG to have Hermeians significantly depends on the MW-M31 distance. \textit{Right panel:} MW analogues mass distribution in LG-like systems with and without Hermeian haloes.}
\label{LG with(no) HH}
\end{figure*}

We found 108 Local Group-like pairs in the ESMDPL simulation. Approximately 14 per cent contain Hermeian haloes, and a total of 59 Hermeians were found in the Local Group-like systems. The systems contain from 1 to 13 Hermeian haloes, with an average of approximately 4 (see the left panel of Fig.~\ref{fig: Number of Hermeian haloes in LG and t_bp}). The study conducted in \cite{Newton} revealed that 121 Hermeian haloes were found in one LG analogue. The numerical resolution of the Hestia simulations is higher than that of ESMDPL, so we expect to find more Hermeians in Hestia. We find only two LG-like systems containing at least 10 Hermeian haloes. However, the population of 121 Hermeians in \cite{Newton} includes all the haloes, including those with less than 100 simulation particles. If we remove the 100 particle limit on the haloes that we consider to be Hermeian, the maximal number of Hermeians in LG-like systems would increase to 56. Counting the number of haloes below this 100 particles limit can result in an underestimation because low mass objects are more susceptible to artificial disruption due to the numerical effects caused by the limited resolution. Relatedly, any conclusions drawn about the internal properties of the haloes with fewer than 100 particles are likely to suffer too. The onset of significant numerical effects could differ between two different simulations, so comparing the number counts directly is not a robust approach. The appearance of the large number of Hermeians in some LG-like systems deserves further investigation.

%Thus, a separate study of such systems is required in order to find out the reasons for the appearance of quite large number of haloes in them.
%In fact, the ESMDPL is a constrained simulation, i.e. it reproduces observed structures, in particular, the known nearby objects: the MW and M31 (i.e. the LG replica in \cite{Sorce:Ocvirk:Aubert:2022}). These haloes are among the targets of Hermeian haloes, 36 Hermeian haloes passed through one of them. However, there are no Hermeians that passed through both the MW and M31 in this LG replica.

According to the assumptions of \cite{Newton}, the presence of Hermeian haloes may be influenced by the masses of the Local Group hosts and their location relative to each other. We find that primary pairs of LG analogues with Hermeian haloes tend to be closer together at $z=0$ than those of LG-like systems without Hermeians (see Fig.~\ref{LG with(no) HH} left panel). This is consistent with the findings of \cite{Newton}, where one HESTIA simulation yields more than 90 per cent of all the Hermeian haloes that were found. That Local Group has the most massive primary hosts that are likewise 20 per cent closer to each other at z = 0 than in the other simulations. However, from the right panel of Fig.~\ref{LG with(no) HH}, one can see that the mass of the MW does not play a crucial role in the presence of Hermeians. Finally, if we restrict the sample of LG analogues to those with a MW-M31 separation smaller than 1~$\mathrm{Mpc}\;h^{-1}$, 15 out of the 49 analogues host Hermeian haloes.
\begin{figure}
\centering\includegraphics[width=\linewidth]{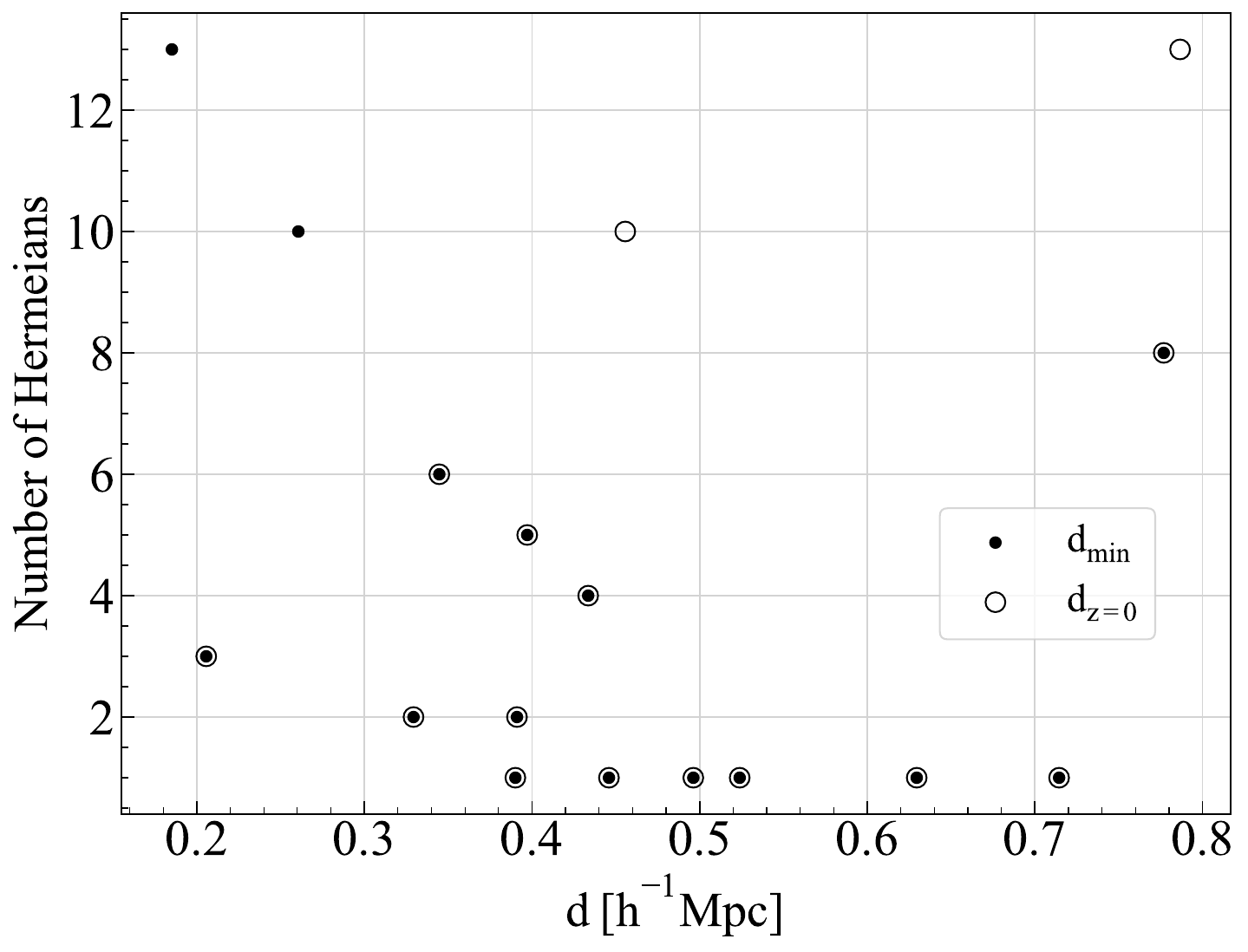}
\caption{Number of Hermeian haloes in a LG analogue as a function of the distance between the MW and M31 analogues.The empty and filled markers show the distance between the host haloes at z = 0, and the minimum distance between them from the moment of the earliest second passage of their Hermeians until z = 0, respectively.}   %The colours indicate the average of the present CIC overdensities at the locations of the MW and M31. The overdensity is given in units of its mean value. To compute the environmental density we implement CIC smoothing with a voxel size of 0.5 $h^{-1}\; \mathrm{Mpc}$ and convolution with HWHM $= 1.6$ $h^{-1}\; \mathrm{Mpc}$ Gaussian smoothing kernel.}
\label{fig: HH_number_dist_Density}
\end{figure}

To better understand what determines the quantity of Hermeian haloes in LG-like pairs, we plot the number of Hermeians as a function of the distance between their targets in Figure~\ref{fig: HH_number_dist_Density}. %and large-scale density.
In some of the selected LG analogues the MW and M31 passed closer to each other in the past than they are at $z=0$, so we consider two distance measures here: the distance at $z=0$, and the minimal comoving separation. One can see that there is no clear trend with the present distance between the MW and M31 analogues. In contrast, the figure shows that systems with smaller lifetime minimum distances between the target haloes typically host more Hermeians. The notable exception to this trend is the system with 8 Hermeians that has a large distance between its hosts (> 0.75 $h^{-1}\; \mathrm{Mpc}$). This system is characterized by a higher environmental density than the other LG analogues with Hermeians. The LG with 8 Hermeians has a $z=0$ CIC overdensity $\rho / \Bar{\rho} = 4.5$, while the average of the overdensities of all LG analogues with Hermeians is 2.5 with a standard deviation of 1.2. This finding suggests that systems where Hermeian haloes are most abundant appear to have closer primary hosts or be located at higher densities. 

%The distribution of the amount of time that Hermeians spend between the MW and M31 is shown in the right panel of Fig.~\ref{fig: Number of Hermeian haloes in LG and t_bp}. One can see that this period is below 7~Gyr for all the Hermeians, but the distribution is quite wide.

57 out of 59 Hermeian haloes found in our LG-like systems have masses below $10^9$~M$_\odot$, so the most of Hermeian haloes in LG-like systems may be devoid of stars due to the photoheating by the UV background. This also has been shown in \cite{Newton} where the photoheating has been taken into account in the simulation. In \cite{Newton}, 97\% (133 out of 137) Hermeian haloes have no stars.

%however, it is possible that these results were influenced by the small sample size.} %However, systems where Hermeian haloes are most abundant appear to have closer primary hosts or be located at higher densities.   
\subsubsection{Spatial distribution}
Like in \cite{Newton}, in order to characterize the spatial distribution of  haloes in a system, we use the angle formed by a considered halo and the M31 analogue with respect to the midpoint of the line connecting the MW and M31 analogues, and we also calculate the distance from a Hermeian halo to the midpoint.
\begin{figure*}
\centering\includegraphics[width=\textwidth]{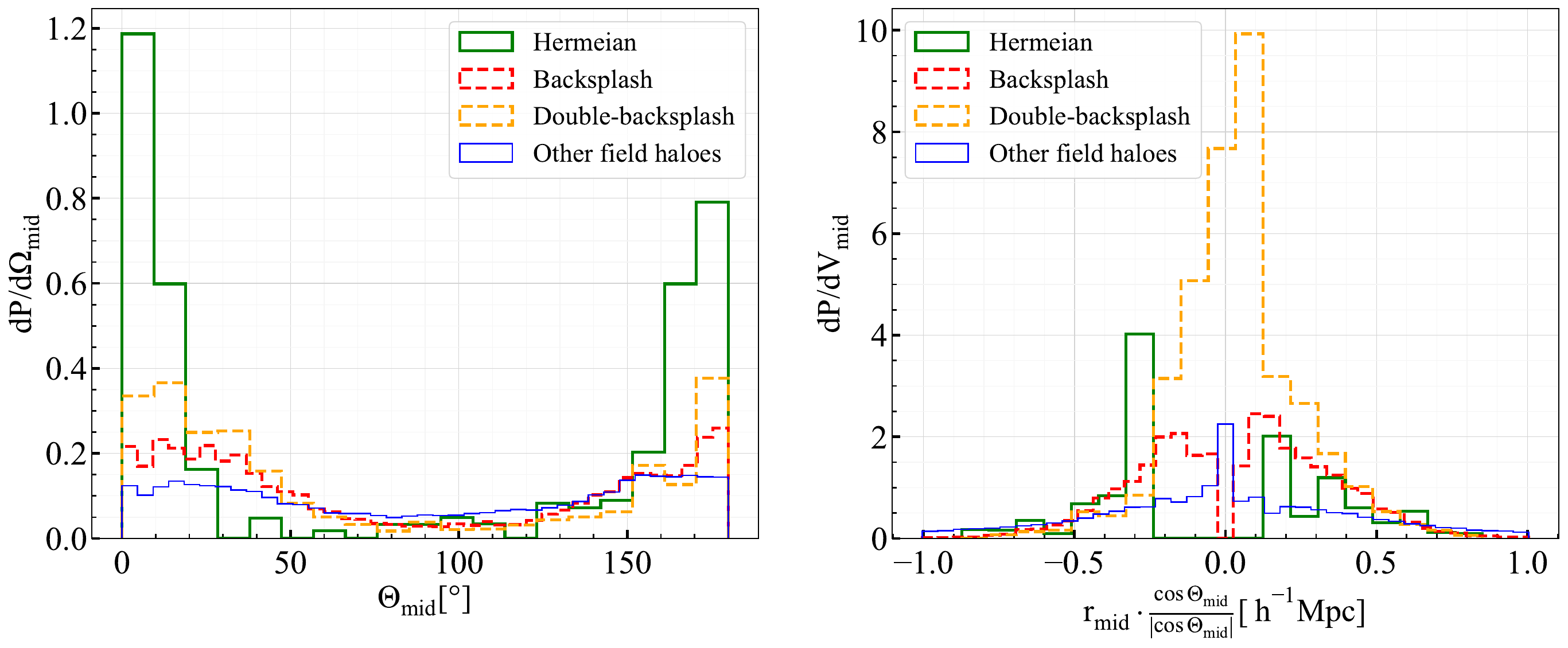}
\caption{\textit{Left panel:} Probability distribution of angles, $\mathrm{\Theta_{mid}}$, that the backsplash, double-backsplash, Hermeian, and remaining field haloes make with the vector towards the M31 analogue from the midpoint of the line connecting the MW and M31 analogues. In this basis, the M31 analogue is at $\mathrm{\Theta_{mid} = 0^{\circ}}$ and the MW analogue is at $\mathrm{\Theta_{mid} = 180^{\circ}}$. \textit{Right panel:} Probability density functions of the distances, $\mathrm{r_{mid}}$, of the haloes from the midpoint of the MW-M31 line. It is multiplied by sgn ($\mathrm{\cos\Theta_{mid}}$), so that haloes in the direction of the M31 analogue have positive values and haloes in the hemisphere containing the MW analogue have negative values. Field haloes were selected within $\sim 1\; h^{-1}\; \mathrm{Mpc}$ from the midpoint of the system (as the radius of the sphere, we use the greatest distance from a Hermeian halo to the midpoint of the line connecting the MW and M31 analogues).}
\label{fig: spatial distribution}
\end{figure*}
The left panel of Figure~\ref{fig: spatial distribution} shows that in contrast to other field haloes, Hermeians are grouped along the line connecting the MW and M31. The backsplash haloes have a comparable angular distribution to the field haloes that have not interacted with the primary hosts of the LG analogues, however, they are more concentrated towards the MW and M31 than other field haloes, excluding Hermeians. It is interesting to note that, similar to \cite{Newton}, we find that a larger fraction of backsplashers are located around the line connecting the MW and M31 analogues in the direction of M31. It can also be seen from the right panel of Fig.~\ref{fig: spatial distribution} that the backsplashers tend to be slightly less concentrated around the LG midpoint than the Hermeians, but are significantly more concentrated in this location than remaining field haloes. Surprisingly, double backsplashers are the most concentrated towards the LG midpoint among the populations we consider. The reason for that is not clear and requires further investigation. These findings should facilitate the identification of Hermeian haloes in observations and confirm the results of \cite{Newton} with greater statistical certainty.

Interestingly, in the LG analogue with the second largest population of Hermeian haloes in our sample (10 haloes), the Hermeians are located perpendicular to the line connecting the MW and M31. This is likely to be the result of the the close passage of the MW and M31 in the past in this particular LG analogue.

\subsubsection{Velocity distribution}
Understanding the differences in the velocity distributions of Hermeian and other field haloes might help to determine whether an observed halo belongs to the Hermeian population. This part of the study is motivated by the presence of a `plane of high-velocity galaxies' in the LG \cite{Banic:Zhao:2017}. There are several galaxies, including the dwarf galaxy NGC 3109, that have much higher radial velocities than predicted by the standard cosmological model. 
The key feature of NGC 3109 is the unexpectedly high radial velocity\footnote{Here we consider the full radial velocity, peculiar + Hubble.} with which it is receding from the MW \cite{Banik:Haslbauer:Pawlowski:2021}. Here we compare the velocity of NGC 3109 with the radial velocities of Hermeian haloes with respect to their second targets.
%Since the key feature of NGC 3109 is the unexpectedly high radial velocity\footnote{Here we consider the full radial velocity, peculiar + Hubble.} with which it is receding from the MW \cite{Banik:Haslbauer:Pawlowski:2021}, here we examine only those Hermeian haloes whose second target is a MW analogue. The MW is chosen as the second target, because the observed high-velocity galaxy NGC 3109 is located in the direction towards the MW, and not M31. %Therefore, one should not study galaxies located on the other side of the M31 for an observer inside the MW (there are Hermeian haloes whose second target is a M31 analogue). 
We select non-Hermeian field haloes for the analysis within 1.5 $h^{-1} \;\mathrm{Mpc}$ spheres centered on the second target, since this is the maximum distance between the Hermeian haloes and their second targets in the LG-like systems that we consider. The study conducted in \cite{Banik:Haslbauer:Pawlowski:2021} has already considered NGC 3109 as a backsplasher candidate, so we adopt the same properties of this galaxy in this work. The observed Galactocentric distance and radial velocity of NGC 3109 are 1.3 Mpc and 170 $\mathrm{km\;s^{-1}}$,  respectively \cite{Banik:Haslbauer:Pawlowski:2021,Soszynski:Gieren:Pietrzy:2006, Dalcanton:Williams:Seth:2009}. Nevertheless, like \cite{ Banik:Haslbauer:Pawlowski:2021}, we require that NGC 3109 analogues should be at least 1.2 Mpc away from their second targets. We normalize the velocity of NGC 3109 and the simulated Hermeian haloes by the circular velocity of the MW halo, to remove the dependency of velocity on the host halo mass.  In order to calculate the circular velocity of the MW in observations, we assume that the MW mass is in the range $[1 - 2.4]\times \mathrm{10^{12}\;M_{\odot}}$ \cite{Boylan_Kolchin_2013}. For the simulated Hermeian haloes we take the mass of the corresponding second target.

\begin{figure}
\centering\includegraphics[width=\linewidth]{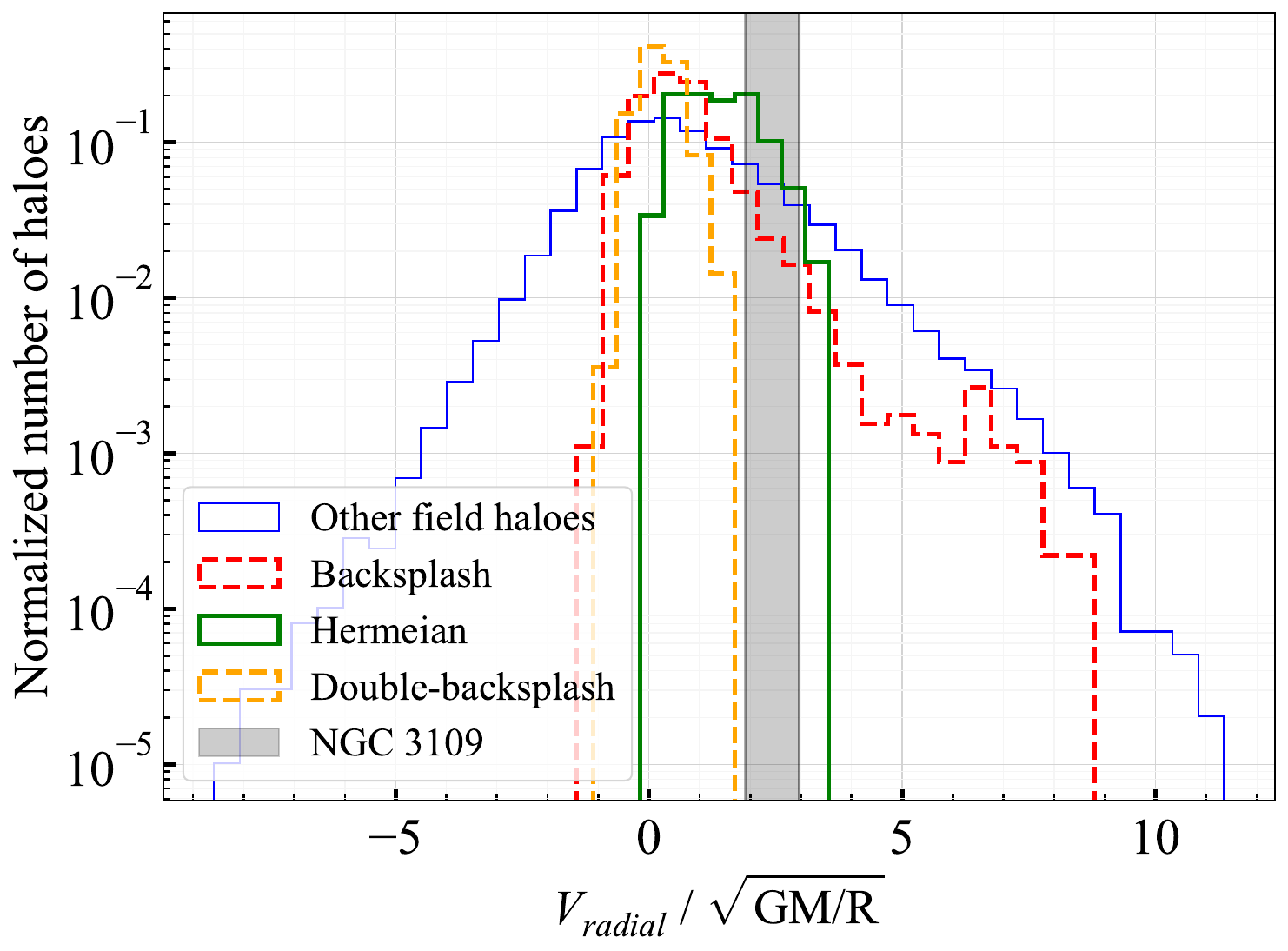}
\caption{Distribution of radial velocities of Hermeian haloes relative to their second targets, backsplash and double-backsplash haloes relative to their targets, and the remaining field haloes relative to the MW and M31 analogues at z = 0. Field haloes were selected within $1.5\; h^{-1}\; \mathrm{Mpc}$ spheres centered on a MW and M31 analogue (as the radius of the sphere, we use the greatest distance from a Hermeian halo to the second target). The shaded region shows the estimated velocity of the NGC 3109 at different MW masses. %\textit{Right panel:} Same as left panel, but on a logarithmic scale.
The halo velocity is given in units of the circular velocity of the corresponding "target halo". Haloes that are moving away from their “target haloes” have positive values of relative velocity and haloes that are moving towards them have negative values.}
\label{fig: velocity}
\end{figure}
%The vast majority of Hermeian haloes that passed through the M31 first and then through the MW are now moving away from their second targets
All Hermeian haloes of the LG analogues are moving away from their second targets at $z=0$, while 64 per cent of other field haloes (excluding backsplashers) are approaching them. The median radial velocity of Hermeians $v_r = 1.4$ in units of $\sqrt{GM/R}$, which is much higher than that of backsplash haloes (0.55), double-backsplash haloes (0.24), and other field haloes (0.5, see Fig.~\ref{fig: velocity}). %According to our preliminary estimates, 
The higher median velocity of Hermeian haloes could be explained by a gravity assist. Moreover, backtracking of the orbits of high-velocity NGC 3109 association galaxies demonstrates that NGC 3109 could have experienced two gravitational assists \cite{Shaya:Tully:2013}. Surprisingly, our estimate shows that, on average, passing through two targets does not lead to a significant change in the halo speed. However, this result does not contradict the possibility that Hermeian haloes gain their high velocities through gravitational assists (because their speeds are not higher than the estimated maximum). It is possible that the increase in velocity due to the gravity assists is compensated by dynamical friction losses during the passage through another halo.

On the other hand, it can be seen in Fig.~\ref{fig: velocity} that the velocity distribution of other field haloes is quite wide. Therefore, an alternative hypothesis is that Hermeians are biased high. In other words, those haloes that initially have high velocities become Hermeians.
The velocity distribution admits that NGC 3109 can be Hermeian, consistent with the suggestion by \cite{Shaya:Tully:2013}   that this galaxy could have experienced two gravity assists.

\subsubsection{Distance to the second targets}
\begin{figure}
\centering\includegraphics[width=\linewidth]{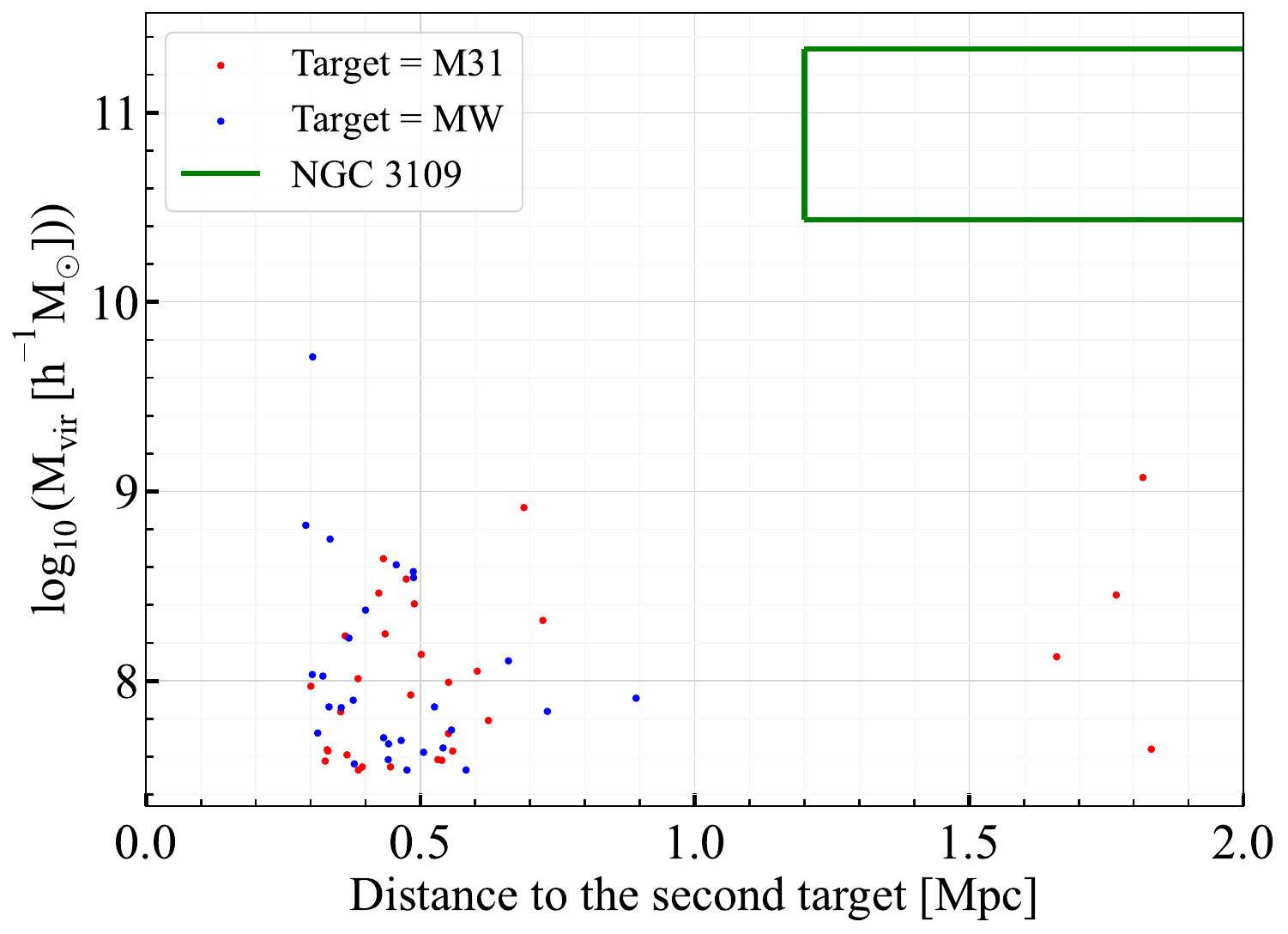}
\caption{Search for Hermeian haloes resembling NGC 3109 in LG-like systems on the plane virial mass --- distance from the second target. The region on the plane corresponding to the estimated mass and distance to NGC 3109 is shown with the green box. Blue and red points are Hermeians which had correspondingly MW and M31 as their second target.}
\label{fig: distance to the second targets}
\end{figure}
The mass-distance distribution Fig.~\ref{fig: distance to the second targets} shows that all Hermeian haloes whose second targets were MW analogues are less massive than NGC 3109 and are located at smaller distances from the second target. At the same time, four Hermeian haloes that first passed through the MW and then through M31 are more than 1.5 Mpc away from their second targets at $z=0$. Therefore, it is sometimes possible to find Hermeians with similar distances as NGC 3109. This is a reassuring finding, since the masses of the host haloes (the MW and M31) in LG-like systems could be close to each other, which means that such different results for them would be due to insufficient sample size. However, the masses of these most distant Hermeian haloes do not exceed $\mathrm{\sim 1.5 \times 10^{9}\;M_{\odot}}$, which is much smaller than the lower boundary of the interval in which the estimated mass of NGC 3109 lies (like \cite{Banik:Haslbauer:Pawlowski:2021}, we take the mass of NGC 3109 in
the range $[4-32] \times \mathrm{10^{10}\;M_{\odot}}$). Thus, the question of whether this galaxy can be classified as a Hermeian one requires further study.

\section{Conclusions} \label{sec: Conclusions}
In this work we study populations of Hermeian haloes in dark matter simulations. We define Hermeian haloes as objects that have been inside the virial radius of two haloes that are distinct at $z=0$ exactly twice. One can find a large variety of evolutionary tracks of haloes in a dark matter simulation, including multiple passages through the same or different haloes. The most abundant and well studied are backsplash haloes, which have experienced only one passage through another halo. Hydrodynamical simulations show that the passage of a galaxy through another halo leaves a noticeable imprint on its properties, resulting in the quenching of star formation. This makes backsplash galaxies an important target in studies of the link between assembly history and stellar properties of galaxies at $z=0$. 

Hermeian haloes, first identified by \cite{Newton} in constrained simulations of the Local Group, have several important features that make them distinct from backsplash haloes and also interesting objects to study as their own right. As has been shown in \cite{Newton}, they have higher dark matter concentrations, and a peculiar spatial distribution with respect to the LG host galaxies. This motivated this study, in which we identify Hermeian haloes and their targets in the ESMDPL simulation. In addition, to test the numerical effects and convergence of our algorithm, we use the VSMDPL simulation that has a lower mass resolution but more frequent snapshots. Our research focuses mainly on the size of the Hermeian halo population and its characteristics, as well as the mass and spatial distribution of their targets. Specifically, LG-like systems containing Hermeians were examined. Or results are calculated for haloes with $\mathrm{M_{vir}} >3.3\times10^7\;h^{-1}$~M$_\odot$, which corresponds to 100 simulation particles and gives consistent results according to our investigation of the numerical effects.

Depending on the mass range, backsplashers make up approximately 4 to 13 per cent of all field haloes at $z = 0$, and up to 40 per cent in overdense regions (see Figs.~\ref{fig: Fractions of haloes}, \ref{fig: Density_revers}). Compared to the size of the backsplash halo population, the Hermeian one is on average 8.5 times smaller.  However, we find that the Hermeian haloes can be ubiquitous, with more than 95 per cent of haloes being targets for at least one Hermeian (see Fig.~\ref{fig: Targets}). We show that the majority of the targets of Hermeian haloes ($\textgreater$ 62 per cent) are no more than $0.4\;h^{-1} \;\mathrm{Mpc}$ apart at z = 0. The mass ratio of the two targets varies uniformly between $10^{-4}$ and $10^3$; however the first target is usually less massive than the second one. Half of the Hermeian haloes have spend more than 2.1~Gyr between passages through the target haloes (see Fig.~\ref{fig:Time between passages all HH}). For most targets (77 per cent) only one Hermeian halo passed through it, nevertheless, we find that galaxy cluster-mass targets that have been passed through several thousand times (see Fig.~\ref{fig: Targets}).

When we considered MW--M31 pairs with distances up to $1.5\;h^{-1} \;\mathrm{Mpc}$, approximately one in seven of the selected analogues of the Local Group contains Hermeian haloes. However, when the distance is limited to $1\;h^{-1}\; \mathrm{Mpc}$, this fraction growths to 30 per cent. If we relax the requirement that Hermeian haloes must have at least 100 simulation particles, as much as 50\% of LG analogues with MW-M31 distance below $1\;h^{-1}\; \mathrm{Mpc}$ host Hermeian population. However, our sample is not complete at these low masses, below 100 particles, so this fraction could be even higher. We find that LG-like systems host on average 4 Hermeians with $\mathrm{M_{vir}>3.3}\times10^7\;h^{-1}$~M$_\odot$. Therefore, multiple passages with hundreds of Hermeian haloes should be considered as rare events (they might be less rare if we could resolve very low-mass haloes). Hermeian haloes in LG-like pairs, as well as in other systems, are approximately twice as concentrated as the remaining field halo population, and 20\% more concentrated than backsplash haloes (see Fig.~\ref{fig:Concentration}). The fraction of Hermeian haloes in dense local environments is up to an order of magnitude higher than on average in the Universe (see Fig.~\ref{fig: Density_revers}). The median overdensity of the environment of the Hermeians is also two times higher than that of backsplashers (see Fig.~\ref{fig: Density}). 

%In other words, in high density environments the fraction of Hermeian haloes becomes closer to that of backsplash haloes.

We find that Hermeian haloes are grouped along the line connecting the MW and M31 analogues, whereas the non-Hermeians are distributed more uniformly. This is consistent with the results of \cite{Newton} and could help to identify them in observations. However it is obviously difficult to observe small dwarf galaxies behind M31 due to the galaxy's high stellar density.  Most of the Hermeians were found at distances of less than 1 Mpc from their second targets, but we also find 4 Hermeian haloes more than 1.6 Mpc away from the second target. Hermeian haloes also have high velocities relative to their second targets, and relative to the surrounding field haloes. Half of all Hermeian haloes move faster than 240~$\mathrm{km\;s^{-1}}$ with respect to haloes within 1 $h^{-1}\; \mathrm{Mpc}$ around them, which is somewhat higher than for backsplash (175~$\mathrm{km\;s^{-1}}$) and double-backsplash (202~$\mathrm{km\;s^{-1}}$) haloes. This supports the hypothesis that the fast-moving galaxy NGC 3109 may be a Hermeian galaxy.

Double-backsplash haloes, like Hermeians, experienced two interactions in the past, but with the same target halo. They have even higher concentrations than Hermeians, but lower velocities with respect to the target and the environment (see Fig.~\ref{fig: velocity}). Double-backsplash haloes are likely to be subhaloes that orbit the target on elongated orbits, so they are gravitationally bound to their targets. Unlike Hermeians, they should be considered as subhaloes rather than field haloes.

In conclusion, our findings show that Hermeian galaxies are an interesting class of objects. They are a prediction of the $\Lambda$CDM cosmological model and are a consequence of the hierarchical galaxy assembly paradigm. They constitute few per cent of the field galaxy population and their properties may allow them to be distinguished from backsplash galaxies and regular field galaxies. Hermeian galaxies deserve deeper study, for example in hydrodynamical simulations. This will allow us to understand how the second passage could affect the properties of a galaxy, and then, probably, such objects could be identified in observations.

\section*{Declaration of Competing Interest}
The authors declare that they have no known competing financial interests or personal relationships that could have appeared to influence the work reported in this paper.

\section*{Acknowledgements}
The authors gratefully acknowledge the Gauss Centre for Supercomputing e.V. (www.gauss-centre.eu) for funding this project by providing computing time on the GCS Supercomputer SUPERMUC-NG at Leibniz Supercomputing Centre (www.lrz.de). We thank Peter Behroozi for creating and providing the ROCKSTAR merger trees of the VSMDPL and ESMDPL simulations. The CosmoSim database (https://www.cosmosim.org) provides access to the simulation and the ROCKSTAR data. The database is a service by the Leibniz Institute for Astrophysics Potsdam (AIP).
JS acknowledges support from the French Agence Nationale de la Recherche for the LOCALIZATION project under grant agreements ANR-21-CE31-001.
The work of SP was supported by LPI NNG 41-2020.
%added by Gyepes
GY also acknowledges the  Ministerio de Ciencia e Innovaci\'on (Spain)  for partial financial support under research grant PID2021-122603NB-C21.
ON is supported by grant 2020/39/B/ST9/03494 from the Polish National Science Centre.

\def\apj{Astrophys.~J}
\def\apjl{Astrophys.~J.,~Lett}
\def\apjs{Astrophys.~J.,~Supplement}
\def\an{Astron.~Nachr}
\def\aap{Astron.~Astrophys}
\def\mnras{Mon.~Not.~R.~Astron.~Soc}
\def\pasp{Publ.~Astron.~Soc.~Pac}
\def\aaps{Astron.~and Astrophys.,~Suppl.~Ser}
\def\apss{Astrophys.~Space.~Sci}
\def\ibvs{Inf.~Bull.~Variable~Stars}
\def\japa{J.~Astrophys.~Astron}
\def\na{New~Astron}
\def\aspproc{Proc.~ASP~conf.~ser.}
\def\aspcs{ASP~Conf.~Ser}
\def\aj{Astron.~J}
\def\actaa{Acta Astron}
\def\araa{Ann.~Rev.~Astron.~Astrophys}
\def\caosp{Contrib.~Astron.~Obs.~Skalnat{\'e}~Pleso}
\def\pasj{Publ.~Astron.~Soc.~Jpn}
\def\memsai{Mem.~Soc.~Astron.~Ital}
\def\astl{Astron.~Letters}
\def\aipproc{Proc.~AIP~conf.~ser.}
\def\physrep{Physics Reports}
\def\jcap{Journal of Cosmology and Astroparticle Physics}
 \bibliographystyle{elsarticle-num} 
 \bibliography{references}

\begin{thebibliography}{10}
\expandafter\ifx\csname url\endcsname\relax
  \def\url#1{\texttt{#1}}\fi
\expandafter\ifx\csname urlprefix\endcsname\relax\def\urlprefix{URL }\fi
\expandafter\ifx\csname href\endcsname\relax
  \def\href#1#2{#2} \def\path#1{#1}\fi

\bibitem{Gill:Knebe:Gibson:2005}
S.~P.~D. {Gill}, A.~{Knebe}, B.~K. {Gibson}, {The evolution of substructure -
  III. The outskirts of clusters}, \mnras 356~(4) (2005) 1327--1332.
\newblock \href {http://arxiv.org/abs/astro-ph/0404427}
  {\path{arXiv:astro-ph/0404427}}, \href
  {https://doi.org/10.1111/j.1365-2966.2004.08562.x}
  {\path{doi:10.1111/j.1365-2966.2004.08562.x}}.

\bibitem{Moore04}
B.~{Moore}, J.~{Diemand}, J.~{Stadel}, {On the age-radius relation and orbital
  history of cluster galaxies}, in: A.~{Diaferio} (Ed.), IAU Colloq. 195:
  Outskirts of Galaxy Clusters: Intense Life in the Suburbs, 2004, pp.
  513--518.
\newblock \href {http://arxiv.org/abs/astro-ph/0406615}
  {\path{arXiv:astro-ph/0406615}}, \href
  {https://doi.org/10.1017/S1743921304001127}
  {\path{doi:10.1017/S1743921304001127}}.

\bibitem{Knebe11a}
A.~{Knebe}, N.~I. {Libeskind}, S.~R. {Knollmann}, L.~A. {Martinez-Vaquero},
  G.~{Yepes}, S.~{Gottl{\"o}ber}, Y.~{Hoffman}, {The luminosities of backsplash
  galaxies in constrained simulations of the Local Group}, \mnras 412~(1)
  (2011) 529--536.
\newblock \href {http://arxiv.org/abs/1010.5670} {\path{arXiv:1010.5670}},
  \href {https://doi.org/10.1111/j.1365-2966.2010.17924.x}
  {\path{doi:10.1111/j.1365-2966.2010.17924.x}}.

\bibitem{An:Kim:Moon:2019}
S.-H. An, J.~Kim, J.-S. Moon, S.-J. Yoon,
  \href{https://doi.org/10.3847/1538-4357/ab535f}{Living with neighbors. {II}.
  statistical analysis of flybys and mergers of dark matter halos in
  cosmological simulations}, The Astrophysical Journal 887~(1) (2019) 59.
\newblock \href {https://doi.org/10.3847/1538-4357/ab535f}
  {\path{doi:10.3847/1538-4357/ab535f}}.
\newline\urlprefix\url{https://doi.org/10.3847/1538-4357/ab535f}

\bibitem{Ludlow09}
A.~D. {Ludlow}, J.~F. {Navarro}, V.~{Springel}, A.~{Jenkins}, C.~S. {Frenk},
  A.~{Helmi}, {The Unorthodox Orbits of Substructure Halos}, \apj 692~(1)
  (2009) 931--941.
\newblock \href {http://arxiv.org/abs/0801.1127} {\path{arXiv:0801.1127}},
  \href {https://doi.org/10.1088/0004-637X/692/1/931}
  {\path{doi:10.1088/0004-637X/692/1/931}}.

\bibitem{Teyssier:Johnston:Kuhlen:2012}
M.~{Teyssier}, K.~V. {Johnston}, M.~{Kuhlen}, {Identifying Local Group field
  galaxies that have interacted with the Milky Way}, \mnras 426~(3) (2012)
  1808--1818.
\newblock \href {http://arxiv.org/abs/1207.2768} {\path{arXiv:1207.2768}},
  \href {https://doi.org/10.1111/j.1365-2966.2012.21793.x}
  {\path{doi:10.1111/j.1365-2966.2012.21793.x}}.

\bibitem{GarrisonKimmel14}
S.~{Garrison-Kimmel}, M.~{Boylan-Kolchin}, J.~S. {Bullock}, K.~{Lee}, {ELVIS:
  Exploring the Local Volume in Simulations}, \mnras 438~(3) (2014) 2578--2596.
\newblock \href {http://arxiv.org/abs/1310.6746} {\path{arXiv:1310.6746}},
  \href {https://doi.org/10.1093/mnras/stt2377}
  {\path{doi:10.1093/mnras/stt2377}}.

\bibitem{Bakels21}
L.~{Bakels}, A.~D. {Ludlow}, C.~{Power}, {Pre-processing, group accretion, and
  the orbital trajectories of associated subhaloes}, \mnras 501~(4) (2021)
  5948--5963.
\newblock \href {http://arxiv.org/abs/2008.05475} {\path{arXiv:2008.05475}},
  \href {https://doi.org/10.1093/mnras/staa3979}
  {\path{doi:10.1093/mnras/staa3979}}.

\bibitem{Green21}
S.~B. {Green}, F.~C. {van den Bosch}, F.~{Jiang}, {The tidal evolution of dark
  matter substructure - II. The impact of artificial disruption on subhalo mass
  functions and radial profiles}, \mnras 503~(3) (2021) 4075--4091.
\newblock \href {http://arxiv.org/abs/2103.01227} {\path{arXiv:2103.01227}},
  \href {https://doi.org/10.1093/mnras/stab696}
  {\path{doi:10.1093/mnras/stab696}}.

\bibitem{Knebe11b}
A.~{Knebe}, N.~I. {Libeskind}, T.~{Doumler}, G.~{Yepes}, S.~{Gottl{\"o}ber},
  Y.~{Hoffman}, {Renegade subhaloes in the Local Group}, \mnras 417~(1) (2011)
  L56--L60.
\newblock \href {http://arxiv.org/abs/1107.2944} {\path{arXiv:1107.2944}},
  \href {https://doi.org/10.1111/j.1745-3933.2011.01119.x}
  {\path{doi:10.1111/j.1745-3933.2011.01119.x}}.

\bibitem{Wetzel15}
A.~R. {Wetzel}, A.~J. {Deason}, S.~{Garrison-Kimmel}, {Satellite Dwarf Galaxies
  in a Hierarchical Universe: Infall Histories, Group Preprocessing, and
  Reionization}, \apj 807~(1) (2015) 49.
\newblock \href {http://arxiv.org/abs/1501.01972} {\path{arXiv:1501.01972}},
  \href {https://doi.org/10.1088/0004-637X/807/1/49}
  {\path{doi:10.1088/0004-637X/807/1/49}}.

\bibitem{vandenBosch17}
F.~C. {van den Bosch}, {Dissecting the evolution of dark matter subhaloes in
  the Bolshoi simulation}, \mnras 468~(1) (2017) 885--909.
\newblock \href {http://arxiv.org/abs/1611.02657} {\path{arXiv:1611.02657}},
  \href {https://doi.org/10.1093/mnras/stx520}
  {\path{doi:10.1093/mnras/stx520}}.

\bibitem{Balogh00}
M.~L. {Balogh}, J.~F. {Navarro}, S.~L. {Morris}, {The Origin of Star Formation
  Gradients in Rich Galaxy Clusters}, \apj 540~(1) (2000) 113--121.
\newblock \href {http://arxiv.org/abs/astro-ph/0004078}
  {\path{arXiv:astro-ph/0004078}}, \href {https://doi.org/10.1086/309323}
  {\path{doi:10.1086/309323}}.

\bibitem{Simpson18}
C.~M. {Simpson}, R.~J.~J. {Grand}, F.~A. {G{\'o}mez}, F.~{Marinacci},
  R.~{Pakmor}, V.~{Springel}, D.~J.~R. {Campbell}, C.~S. {Frenk}, {Quenching
  and ram pressure stripping of simulated Milky Way satellite galaxies}, \mnras
  478~(1) (2018) 548--567.
\newblock \href {http://arxiv.org/abs/1705.03018} {\path{arXiv:1705.03018}},
  \href {https://doi.org/10.1093/mnras/sty774}
  {\path{doi:10.1093/mnras/sty774}}.

\bibitem{Buck19}
T.~{Buck}, A.~V. {Macci{\`o}}, A.~A. {Dutton}, A.~{Obreja}, J.~{Frings}, {NIHAO
  XV: the environmental impact of the host galaxy on galactic satellite and
  field dwarf galaxies}, \mnras 483~(1) (2019) 1314--1341.
\newblock \href {http://arxiv.org/abs/1804.04667} {\path{arXiv:1804.04667}},
  \href {https://doi.org/10.1093/mnras/sty2913}
  {\path{doi:10.1093/mnras/sty2913}}.

\bibitem{Benavides21}
J.~A. {Benavides}, L.~V. {Sales}, M.~G. {Abadi}, A.~{Pillepich}, D.~{Nelson},
  F.~{Marinacci}, M.~{Cooper}, R.~{Pakmor}, P.~{Torrey}, M.~{Vogelsberger},
  L.~{Hernquist}, {Quiescent ultra-diffuse galaxies in the field originating
  from backsplash orbits}, Nature Astronomy 5 (2021) 1255--1260.
\newblock \href {http://arxiv.org/abs/2109.01677} {\path{arXiv:2109.01677}},
  \href {https://doi.org/10.1038/s41550-021-01458-1}
  {\path{doi:10.1038/s41550-021-01458-1}}.

\bibitem{Joshi21}
G.~D. {Joshi}, A.~{Pillepich}, D.~{Nelson}, E.~{Zinger}, F.~{Marinacci},
  V.~{Springel}, M.~{Vogelsberger}, L.~{Hernquist}, {The cumulative star
  formation histories of dwarf galaxies with TNG50. I: environment-driven
  diversity and connection to quenching}, \mnras 508~(2) (2021) 1652--1674.
\newblock \href {http://arxiv.org/abs/2101.12226} {\path{arXiv:2101.12226}},
  \href {https://doi.org/10.1093/mnras/stab2573}
  {\path{doi:10.1093/mnras/stab2573}}.

\bibitem{Casey22}
K.~J. {Casey}, J.~P. {Greco}, A.~H.~G. {Peter}, A.~B. {Davis}, {Discovery of a
  red backsplash galaxy candidate near M81}, arXiv e-prints (2022)
  arXiv:2211.00629\href {http://arxiv.org/abs/2211.00629}
  {\path{arXiv:2211.00629}}.

\bibitem{Newton}
O.~{Newton}, N.~I. {Libeskind}, A.~{Knebe}, M.~A. {S{\'a}nchez-Conde}, J.~G.
  {Sorce}, S.~{Pilipenko}, M.~{Steinmetz}, R.~{Pakmor}, E.~{Tempel},
  Y.~{Hoffman}, M.~{Vogelsberger}, {Hermeian haloes: Field haloes that
  interacted with both the Milky Way and M31}, \mnras 514~(3) (2022)
  3612--3625.
\newblock \href {http://arxiv.org/abs/2104.11242} {\path{arXiv:2104.11242}},
  \href {https://doi.org/10.1093/mnras/stac1316}
  {\path{doi:10.1093/mnras/stac1316}}.

\bibitem{Libeskind2020}
N.~I. {Libeskind}, E.~{Carlesi}, R.~J.~J. {Grand}, A.~{Khalatyan}, A.~{Knebe},
  R.~{Pakmor}, S.~{Pilipenko}, M.~S. {Pawlowski}, M.~{Sparre}, E.~{Tempel},
  P.~{Wang}, H.~M. {Courtois}, S.~{Gottl{\"o}ber}, Y.~{Hoffman}, I.~{Minchev},
  C.~{Pfrommer}, J.~G. {Sorce}, V.~{Springel}, M.~{Steinmetz}, R.~B. {Tully},
  M.~{Vogelsberger}, G.~{Yepes}, {The HESTIA project: simulations of the Local
  Group}, \mnras 498~(2) (2020) 2968--2983.
\newblock \href {http://arxiv.org/abs/2008.04926} {\path{arXiv:2008.04926}},
  \href {https://doi.org/10.1093/mnras/staa2541}
  {\path{doi:10.1093/mnras/staa2541}}.

\bibitem{Gaskins:2016}
J.~M. {Gaskins}, {A review of indirect searches for particle dark matter},
  Contemporary Physics 57~(4) (2016) 496--525.
\newblock \href {http://arxiv.org/abs/1604.00014} {\path{arXiv:1604.00014}},
  \href {https://doi.org/10.1080/00107514.2016.1175160}
  {\path{doi:10.1080/00107514.2016.1175160}}.

\bibitem{Banic:Zhao:2017}
I.~Banik, H.~Zhao, \href{https://doi.org/10.1093/mnras/stx2596}{{A plane of
  high-velocity galaxies across the Local Group}}, Monthly Notices of the Royal
  Astronomical Society 473~(3) (2017) 4033--4054.
\newblock \href
  {http://arxiv.org/abs/https://academic.oup.com/mnras/article-pdf/473/3/4033/21841707/stx2596.pdf}
  {\path{arXiv:https://academic.oup.com/mnras/article-pdf/473/3/4033/21841707/stx2596.pdf}},
  \href {https://doi.org/10.1093/mnras/stx2596}
  {\path{doi:10.1093/mnras/stx2596}}.
\newline\urlprefix\url{https://doi.org/10.1093/mnras/stx2596}

\bibitem{Banik:Haslbauer:Pawlowski:2021}
I.~Banik, M.~Haslbauer, M.~S. Pawlowski, B.~Famaey, P.~Kroupa,
  \href{https://doi.org/10.1093/mnras/stab751}{{On the absence of backsplash
  analogues to NGC 3109 in the $\Lambda$CDM framework}}, Monthly Notices of the
  Royal Astronomical Society 503~(4) (2021) 6170--6186.
\newblock \href
  {http://arxiv.org/abs/https://academic.oup.com/mnras/article-pdf/503/4/6170/38118573/stab751.pdf}
  {\path{arXiv:https://academic.oup.com/mnras/article-pdf/503/4/6170/38118573/stab751.pdf}},
  \href {https://doi.org/10.1093/mnras/stab751}
  {\path{doi:10.1093/mnras/stab751}}.
\newline\urlprefix\url{https://doi.org/10.1093/mnras/stab751}

\bibitem{Springel}
V.~{Springel}, {The cosmological simulation code GADGET-2}, \mnras 364~(4)
  (2005) 1105--1134.
\newblock \href {http://arxiv.org/abs/astro-ph/0505010}
  {\path{arXiv:astro-ph/0505010}}, \href
  {https://doi.org/10.1111/j.1365-2966.2005.09655.x}
  {\path{doi:10.1111/j.1365-2966.2005.09655.x}}.

\bibitem{Planck2015}
{Planck Collaboration}, P.~A.~R. {Ade}, N.~{Aghanim}, M.~{Arnaud},
  M.~{Ashdown}, J.~{Aumont}, C.~{Baccigalupi}, A.~J. {Banday}, R.~B.
  {Barreiro}, J.~G. {Bartlett}, N.~{Bartolo}, E.~{Battaner}, R.~{Battye},
  K.~{Benabed}, A.~{Beno{\^\i}t}, A.~{Benoit-L{\'e}vy}, J.~P. {Bernard},
  M.~{Bersanelli}, P.~{Bielewicz}, J.~J. {Bock}, A.~{Bonaldi}, L.~{Bonavera},
  J.~R. {Bond}, J.~{Borrill}, F.~R. {Bouchet}, F.~{Boulanger}, M.~{Bucher},
  C.~{Burigana}, R.~C. {Butler}, E.~{Calabrese}, J.~F. {Cardoso},
  A.~{Catalano}, A.~{Challinor}, A.~{Chamballu}, R.~R. {Chary}, H.~C. {Chiang},
  J.~{Chluba}, P.~R. {Christensen}, S.~{Church}, D.~L. {Clements},
  S.~{Colombi}, L.~P.~L. {Colombo}, C.~{Combet}, A.~{Coulais}, B.~P. {Crill},
  A.~{Curto}, F.~{Cuttaia}, L.~{Danese}, R.~D. {Davies}, R.~J. {Davis}, P.~{de
  Bernardis}, A.~{de Rosa}, G.~{de Zotti}, J.~{Delabrouille}, F.~X.
  {D{\'e}sert}, E.~{Di Valentino}, C.~{Dickinson}, J.~M. {Diego}, K.~{Dolag},
  H.~{Dole}, S.~{Donzelli}, O.~{Dor{\'e}}, M.~{Douspis}, A.~{Ducout},
  J.~{Dunkley}, X.~{Dupac}, G.~{Efstathiou}, F.~{Elsner}, T.~A. {En{\ss}lin},
  H.~K. {Eriksen}, M.~{Farhang}, J.~{Fergusson}, F.~{Finelli}, O.~{Forni},
  M.~{Frailis}, A.~A. {Fraisse}, E.~{Franceschi}, A.~{Frejsel}, S.~{Galeotta},
  S.~{Galli}, K.~{Ganga}, C.~{Gauthier}, M.~{Gerbino}, T.~{Ghosh}, M.~{Giard},
  Y.~{Giraud-H{\'e}raud}, E.~{Giusarma}, E.~{Gjerl{\o}w},
  J.~{Gonz{\'a}lez-Nuevo}, K.~M. {G{\'o}rski}, S.~{Gratton}, A.~{Gregorio},
  A.~{Gruppuso}, J.~E. {Gudmundsson}, J.~{Hamann}, F.~K. {Hansen}, D.~{Hanson},
  D.~L. {Harrison}, G.~{Helou}, S.~{Henrot-Versill{\'e}},
  C.~{Hern{\'a}ndez-Monteagudo}, D.~{Herranz}, S.~R. {Hildebrandt}, E.~{Hivon},
  M.~{Hobson}, W.~A. {Holmes}, A.~{Hornstrup}, W.~{Hovest}, Z.~{Huang}, K.~M.
  {Huffenberger}, G.~{Hurier}, A.~H. {Jaffe}, T.~R. {Jaffe}, W.~C. {Jones},
  M.~{Juvela}, E.~{Keih{\"a}nen}, R.~{Keskitalo}, T.~S. {Kisner}, R.~{Kneissl},
  J.~{Knoche}, L.~{Knox}, M.~{Kunz}, H.~{Kurki-Suonio}, G.~{Lagache},
  A.~{L{\"a}hteenm{\"a}ki}, J.~M. {Lamarre}, A.~{Lasenby}, M.~{Lattanzi}, C.~R.
  {Lawrence}, J.~P. {Leahy}, R.~{Leonardi}, J.~{Lesgourgues}, F.~{Levrier},
  A.~{Lewis}, M.~{Liguori}, P.~B. {Lilje}, M.~{Linden-V{\o}rnle},
  M.~{L{\'o}pez-Caniego}, P.~M. {Lubin}, J.~F. {Mac{\'\i}as-P{\'e}rez},
  G.~{Maggio}, D.~{Maino}, N.~{Mandolesi}, A.~{Mangilli}, A.~{Marchini},
  M.~{Maris}, P.~G. {Martin}, M.~{Martinelli}, E.~{Mart{\'\i}nez-Gonz{\'a}lez},
  S.~{Masi}, S.~{Matarrese}, P.~{McGehee}, P.~R. {Meinhold}, A.~{Melchiorri},
  J.~B. {Melin}, L.~{Mendes}, A.~{Mennella}, M.~{Migliaccio}, M.~{Millea},
  S.~{Mitra}, M.~A. {Miville-Desch{\^e}nes}, A.~{Moneti}, L.~{Montier},
  G.~{Morgante}, D.~{Mortlock}, A.~{Moss}, D.~{Munshi}, J.~A. {Murphy},
  P.~{Naselsky}, F.~{Nati}, P.~{Natoli}, C.~B. {Netterfield}, H.~U.
  {N{\o}rgaard-Nielsen}, F.~{Noviello}, D.~{Novikov}, I.~{Novikov}, C.~A.
  {Oxborrow}, F.~{Paci}, L.~{Pagano}, F.~{Pajot}, R.~{Paladini}, D.~{Paoletti},
  B.~{Partridge}, F.~{Pasian}, G.~{Patanchon}, T.~J. {Pearson}, O.~{Perdereau},
  L.~{Perotto}, F.~{Perrotta}, V.~{Pettorino}, F.~{Piacentini}, M.~{Piat},
  E.~{Pierpaoli}, D.~{Pietrobon}, S.~{Plaszczynski}, E.~{Pointecouteau},
  G.~{Polenta}, L.~{Popa}, G.~W. {Pratt}, G.~{Pr{\'e}zeau}, S.~{Prunet}, J.~L.
  {Puget}, J.~P. {Rachen}, W.~T. {Reach}, R.~{Rebolo}, M.~{Reinecke},
  M.~{Remazeilles}, C.~{Renault}, A.~{Renzi}, I.~{Ristorcelli}, G.~{Rocha},
  C.~{Rosset}, M.~{Rossetti}, G.~{Roudier}, B.~{Rouill{\'e} d'Orfeuil},
  M.~{Rowan-Robinson}, J.~A. {Rubi{\~n}o-Mart{\'\i}n}, B.~{Rusholme},
  N.~{Said}, V.~{Salvatelli}, L.~{Salvati}, M.~{Sandri}, D.~{Santos},
  M.~{Savelainen}, G.~{Savini}, D.~{Scott}, M.~D. {Seiffert}, P.~{Serra},
  E.~P.~S. {Shellard}, L.~D. {Spencer}, M.~{Spinelli}, V.~{Stolyarov},
  R.~{Stompor}, R.~{Sudiwala}, R.~{Sunyaev}, D.~{Sutton}, A.~S. {Suur-Uski},
  J.~F. {Sygnet}, J.~A. {Tauber}, L.~{Terenzi}, L.~{Toffolatti}, M.~{Tomasi},
  M.~{Tristram}, T.~{Trombetti}, M.~{Tucci}, J.~{Tuovinen}, M.~{T{\"u}rler},
  G.~{Umana}, L.~{Valenziano}, J.~{Valiviita}, F.~{Van Tent}, P.~{Vielva},
  F.~{Villa}, L.~A. {Wade}, B.~D. {Wandelt}, I.~K. {Wehus}, M.~{White},
  S.~D.~M. {White}, A.~{Wilkinson}, D.~{Yvon}, A.~{Zacchei}, A.~{Zonca},
  {Planck 2015 results. XIII. Cosmological parameters}, \aap 594 (2016) A13.
\newblock \href {http://arxiv.org/abs/1502.01589} {\path{arXiv:1502.01589}},
  \href {https://doi.org/10.1051/0004-6361/201525830}
  {\path{doi:10.1051/0004-6361/201525830}}.

\bibitem{Behroozi:Wechsler:Wu:2013:763:18}
P.~S. {Behroozi}, R.~H. {Wechsler}, H.-Y. {Wu}, M.~T. {Busha}, A.~A. {Klypin},
  J.~R. {Primack}, {Gravitationally Consistent Halo Catalogs and Merger Trees
  for Precision Cosmology}, \apj 763~(1) (2013) 18.
\newblock \href {http://arxiv.org/abs/1110.4370} {\path{arXiv:1110.4370}},
  \href {https://doi.org/10.1088/0004-637X/763/1/18}
  {\path{doi:10.1088/0004-637X/763/1/18}}.

\bibitem{Behroozi:Wechsler:Wu:762:2}
P.~S. {Behroozi}, R.~H. {Wechsler}, H.-Y. {Wu}, {The ROCKSTAR Phase-space
  Temporal Halo Finder and the Velocity Offsets of Cluster Cores}, \apj 762~(2)
  (2013) 109.
\newblock \href {http://arxiv.org/abs/1110.4372} {\path{arXiv:1110.4372}},
  \href {https://doi.org/10.1088/0004-637X/762/2/109}
  {\path{doi:10.1088/0004-637X/762/2/109}}.

\bibitem{Bryan:Norman:1998}
G.~L. {Bryan}, M.~L. {Norman}, {Statistical Properties of X-Ray Clusters:
  Analytic and Numerical Comparisons}, \apj 495~(1) (1998) 80--99.
\newblock \href {http://arxiv.org/abs/astro-ph/9710107}
  {\path{arXiv:astro-ph/9710107}}, \href {https://doi.org/10.1086/305262}
  {\path{doi:10.1086/305262}}.

\bibitem{Rodrigues-Puebla}
A.~{Rodr{\'\i}guez-Puebla}, J.~R. {Primack}, V.~{Avila-Reese}, S.~M. {Faber},
  {Constraining the galaxy-halo connection over the last 13.3 Gyr: star
  formation histories, galaxy mergers and structural properties}, \mnras
  470~(1) (2017) 651--687.
\newblock \href {http://arxiv.org/abs/1703.04542} {\path{arXiv:1703.04542}},
  \href {https://doi.org/10.1093/mnras/stx1172}
  {\path{doi:10.1093/mnras/stx1172}}.

\bibitem{Knebe:Knollman:Muldrew:2011}
A.~Knebe, S.~Knollmann, S.~Muldrew, F.~Pearce, M.~Aragon, Y.~Ascasibar,
  P.~Behroozi, D.~Ceverino, S.~Colombi, J.~Diemand, K.~Dolag, B.~Falck,
  P.~Fasel, J.~Gardner, S.~Gottloeber, C.-H. Hsu, F.~Iannuzzi, A.~Klypin,
  Z.~Lukic, M.~Zemp, Haloes gone mad: The halo-finder comparison project,
  Monthly Notices of the Royal Astronomical Society 415 (2011) 2293.
\newblock \href {https://doi.org/10.1111/j.1365-2966.2011.18858.x}
  {\path{doi:10.1111/j.1365-2966.2011.18858.x}}.

\bibitem{Diemer_2021}
B.~Diemer, \href{https://doi.org/10.3847/1538-4357/abd947}{Flybys, orbits,
  splashback: Subhalos and the importance of the halo boundary}, The
  Astrophysical Journal 909~(2) (2021) 112.
\newblock \href {https://doi.org/10.3847/1538-4357/abd947}
  {\path{doi:10.3847/1538-4357/abd947}}.
\newline\urlprefix\url{https://doi.org/10.3847/1538-4357/abd947}

\bibitem{Haggar:Gray:Pearce:2020}
R.~{Haggar}, M.~E. {Gray}, F.~R. {Pearce}, A.~{Knebe}, W.~{Cui},
  R.~{Mostoghiu}, G.~{Yepes}, {The Three Hundred project: backsplash galaxies
  in simulations of clusters}, \mnras 492~(4) (2020) 6074--6085.
\newblock \href {http://arxiv.org/abs/2001.11518} {\path{arXiv:2001.11518}},
  \href {https://doi.org/10.1093/mnras/staa273}
  {\path{doi:10.1093/mnras/staa273}}.

\bibitem{Mansfield:Kravtsov:2020}
P.~Mansfield, A.~V. Kravtsov, \href{https://doi.org/10.1093/mnras/staa430}{{The
  three causes of low-mass assembly bias}}, Monthly Notices of the Royal
  Astronomical Society 493~(4) (2020) 4763--4782.
\newblock \href
  {http://arxiv.org/abs/https://academic.oup.com/mnras/article-pdf/493/4/4763/32930684/staa430.pdf}
  {\path{arXiv:https://academic.oup.com/mnras/article-pdf/493/4/4763/32930684/staa430.pdf}},
  \href {https://doi.org/10.1093/mnras/staa430}
  {\path{doi:10.1093/mnras/staa430}}.
\newline\urlprefix\url{https://doi.org/10.1093/mnras/staa430}

\bibitem{van_den_Bosch2017}
F.~C. van~den Bosch, \href{https://doi.org/10.1093/mnras/stx520}{{Dissecting
  the evolution of dark matter subhaloes in the Bolshoi simulation}}, Monthly
  Notices of the Royal Astronomical Society 468~(1) (2017) 885--909.
\newblock \href
  {http://arxiv.org/abs/https://academic.oup.com/mnras/article-pdf/468/1/885/11091980/stx520.pdf}
  {\path{arXiv:https://academic.oup.com/mnras/article-pdf/468/1/885/11091980/stx520.pdf}},
  \href {https://doi.org/10.1093/mnras/stx520}
  {\path{doi:10.1093/mnras/stx520}}.
\newline\urlprefix\url{https://doi.org/10.1093/mnras/stx520}

\bibitem{van_den_Bosch:Ogiya2018}
F.~C. van~den Bosch, G.~Ogiya,
  \href{https://doi.org/10.1093/mnras/sty084}{{Dark matter substructure in
  numerical simulations: a tale of discreteness noise, runaway instabilities,
  and artificial disruption}}, Monthly Notices of the Royal Astronomical
  Society 475~(3) (2018) 4066--4087.
\newblock \href
  {http://arxiv.org/abs/https://academic.oup.com/mnras/article-pdf/475/3/4066/23965833/sty084.pdf}
  {\path{arXiv:https://academic.oup.com/mnras/article-pdf/475/3/4066/23965833/sty084.pdf}},
  \href {https://doi.org/10.1093/mnras/sty084}
  {\path{doi:10.1093/mnras/sty084}}.
\newline\urlprefix\url{https://doi.org/10.1093/mnras/sty084}

\bibitem{Newton:Cautun:Jenkins:2018}
O.~Newton, M.~Cautun, A.~Jenkins, C.~S. Frenk, J.~C. Helly,
  \href{https://doi.org/10.1093/mnras/sty1085}{{The total satellite population
  of the Milky Way}}, Monthly Notices of the Royal Astronomical Society 479~(3)
  (2018) 2853--2870.
\newblock \href
  {http://arxiv.org/abs/https://academic.oup.com/mnras/article-pdf/479/3/2853/25149561/sty1085.pdf}
  {\path{arXiv:https://academic.oup.com/mnras/article-pdf/479/3/2853/25149561/sty1085.pdf}},
  \href {https://doi.org/10.1093/mnras/sty1085}
  {\path{doi:10.1093/mnras/sty1085}}.
\newline\urlprefix\url{https://doi.org/10.1093/mnras/sty1085}

\bibitem{Sung-Ho}
S.-H. {An}, J.~{Kim}, J.-S. {Moon}, S.-J. {Yoon}, {Living with Neighbors. II.
  Statistical Analysis of Flybys and Mergers of Dark Matter Halos in
  Cosmological Simulations}, \apj 887~(1) (2019) 59.
\newblock \href {http://arxiv.org/abs/1911.11782} {\path{arXiv:1911.11782}},
  \href {https://doi.org/10.3847/1538-4357/ab535f}
  {\path{doi:10.3847/1538-4357/ab535f}}.

\bibitem{SánchezConde:Prada:2014}
M.~A. Sánchez-Conde, F.~Prada,
  \href{https://doi.org/10.1093/mnras/stu1014}{{The flattening of the
  concentration–mass relation towards low halo masses and its implications
  for the annihilation signal boost}}, Monthly Notices of the Royal
  Astronomical Society 442~(3) (2014) 2271--2277.
\newblock \href
  {http://arxiv.org/abs/https://academic.oup.com/mnras/article-pdf/442/3/2271/3498708/stu1014.pdf}
  {\path{arXiv:https://academic.oup.com/mnras/article-pdf/442/3/2271/3498708/stu1014.pdf}},
  \href {https://doi.org/10.1093/mnras/stu1014}
  {\path{doi:10.1093/mnras/stu1014}}.
\newline\urlprefix\url{https://doi.org/10.1093/mnras/stu1014}

\bibitem{Evans04}
N.~W. Evans, F.~Ferrer, S.~Sarkar,
  \href{https://link.aps.org/doi/10.1103/PhysRevD.69.123501}{A travel guide to
  the dark matter annihilation signal}, Phys. Rev. D 69 (2004) 123501.
\newblock \href {https://doi.org/10.1103/PhysRevD.69.123501}
  {\path{doi:10.1103/PhysRevD.69.123501}}.
\newline\urlprefix\url{https://link.aps.org/doi/10.1103/PhysRevD.69.123501}

\bibitem{NFW1996}
J.~F. {Navarro}, C.~S. {Frenk}, S.~D.~M. {White}, {The Structure of Cold Dark
  Matter Halos}, \apj 462 (1996) 563.
\newblock \href {http://arxiv.org/abs/astro-ph/9508025}
  {\path{arXiv:astro-ph/9508025}}, \href {https://doi.org/10.1086/177173}
  {\path{doi:10.1086/177173}}.

\bibitem{Einasto1965}
J.~{Einasto}, {On the Construction of a Composite Model for the Galaxy and on
  the Determination of the System of Galactic Parameters}, Trudy
  Astrofizicheskogo Instituta Alma-Ata 5 (1965) 87--100.

\bibitem{Navarro_Hayashi_2004}
J.~F. {Navarro}, E.~{Hayashi}, C.~{Power}, A.~R. {Jenkins}, C.~S. {Frenk},
  S.~D.~M. {White}, V.~{Springel}, J.~{Stadel}, T.~R. {Quinn}, {The inner
  structure of {\ensuremath{\Lambda}}CDM haloes - III. Universality and
  asymptotic slopes}, \mnras 349~(3) (2004) 1039--1051.
\newblock \href {http://arxiv.org/abs/astro-ph/0311231}
  {\path{arXiv:astro-ph/0311231}}, \href
  {https://doi.org/10.1111/j.1365-2966.2004.07586.x}
  {\path{doi:10.1111/j.1365-2966.2004.07586.x}}.

\bibitem{Gao:Liang:2008}
L.~{Gao}, J.~F. {Navarro}, S.~{Cole}, C.~S. {Frenk}, S.~D.~M. {White},
  V.~{Springel}, A.~{Jenkins}, A.~F. {Neto}, {The redshift dependence of the
  structure of massive {\ensuremath{\Lambda}} cold dark matter haloes}, \mnras
  387~(2) (2008) 536--544.
\newblock \href {http://arxiv.org/abs/0711.0746} {\path{arXiv:0711.0746}},
  \href {https://doi.org/10.1111/j.1365-2966.2008.13277.x}
  {\path{doi:10.1111/j.1365-2966.2008.13277.x}}.

\bibitem{Navarro:Ludow:2010}
J.~F. {Navarro}, A.~{Ludlow}, V.~{Springel}, J.~{Wang}, M.~{Vogelsberger},
  S.~D.~M. {White}, A.~{Jenkins}, C.~S. {Frenk}, A.~{Helmi}, {The diversity and
  similarity of simulated cold dark matter haloes}, \mnras 402~(1) (2010)
  21--34.
\newblock \href {http://arxiv.org/abs/0810.1522} {\path{arXiv:0810.1522}},
  \href {https://doi.org/10.1111/j.1365-2966.2009.15878.x}
  {\path{doi:10.1111/j.1365-2966.2009.15878.x}}.

\bibitem{Dutton:Maccio:2014}
A.~A. Dutton, A.~V. Macciò, \href{https://doi.org/10.1093/mnras/stu742}{{Cold
  dark matter haloes in the Planck era: evolution of structural parameters for
  Einasto and NFW profiles}}, Monthly Notices of the Royal Astronomical Society
  441~(4) (2014) 3359--3374.
\newblock \href
  {http://arxiv.org/abs/https://academic.oup.com/mnras/article-pdf/441/4/3359/4043254/stu742.pdf}
  {\path{arXiv:https://academic.oup.com/mnras/article-pdf/441/4/3359/4043254/stu742.pdf}},
  \href {https://doi.org/10.1093/mnras/stu742}
  {\path{doi:10.1093/mnras/stu742}}.
\newline\urlprefix\url{https://doi.org/10.1093/mnras/stu742}

\bibitem{Klypin:Yeps:Gottlober:2016}
A.~{Klypin}, G.~{Yepes}, S.~{Gottl{\"o}ber}, F.~{Prada}, S.~{He{\ss}},
  {MultiDark simulations: the story of dark matter halo concentrations and
  density profiles}, \mnras 457~(4) (2016) 4340--4359.
\newblock \href {http://arxiv.org/abs/1411.4001} {\path{arXiv:1411.4001}},
  \href {https://doi.org/10.1093/mnras/stw248}
  {\path{doi:10.1093/mnras/stw248}}.

\bibitem{Prada:Klypin:2012}
F.~{Prada}, A.~A. {Klypin}, A.~J. {Cuesta}, J.~E. {Betancort-Rijo},
  J.~{Primack}, {Halo concentrations in the standard {\ensuremath{\Lambda}}
  cold dark matter cosmology}, \mnras 423~(4) (2012) 3018--3030.
\newblock \href {http://arxiv.org/abs/1104.5130} {\path{arXiv:1104.5130}},
  \href {https://doi.org/10.1111/j.1365-2966.2012.21007.x}
  {\path{doi:10.1111/j.1365-2966.2012.21007.x}}.

\bibitem{Klypin:Trujillo-Gomez:2011}
A.~A. {Klypin}, S.~{Trujillo-Gomez}, J.~{Primack}, {Dark Matter Halos in the
  Standard Cosmological Model: Results from the Bolshoi Simulation}, \apj
  740~(2) (2011) 102.
\newblock \href {http://arxiv.org/abs/1002.3660} {\path{arXiv:1002.3660}},
  \href {https://doi.org/10.1088/0004-637X/740/2/102}
  {\path{doi:10.1088/0004-637X/740/2/102}}.

\bibitem{Pilipenko:Sanchez-Conde:2017}
S.~V. {Pilipenko}, M.~A. {S{\'a}nchez-Conde}, F.~{Prada}, G.~{Yepes}, {Pushing
  down the low-mass halo concentration frontier with the Lomonosov cosmological
  simulations}, \mnras 472~(4) (2017) 4918--4927.
\newblock \href {http://arxiv.org/abs/1703.06012} {\path{arXiv:1703.06012}},
  \href {https://doi.org/10.1093/mnras/stx2319}
  {\path{doi:10.1093/mnras/stx2319}}.

\bibitem{L'Huiller:Park:Kim:2015}
B.~L'Huillier, C.~Park, J.~Kim, The ecology of dark matter haloes i: The rates
  and types of halo interactions, Monthly Notices of the Royal Astronomical
  Society 451 (2015) 5046--5057.
\newblock \href {https://doi.org/10.1093/mnras/stv995}
  {\path{doi:10.1093/mnras/stv995}}.

\bibitem{Muldrew:Croton:Skibba:2011}
S.~I. Muldrew, D.~J. Croton, R.~A. Skibba, F.~R. Pearce, H.~B. Ann, I.~K.
  Baldry, S.~Brough, Y.-Y. Choi, C.~J. Conselice, N.~B. Cowan, A.~Gallazzi,
  M.~E. Gray, R.~Grützbauch, I.-H. Li, C.~Park, S.~V. Pilipenko, B.~J.
  Podgorzec, A.~S.~G. Robotham, D.~J. Wilman, X.~Yang, Y.~Zhang, S.~Zibetti,
  \href{https://doi.org/10.11112Fj.1365-2966.2011.19922.x}{Measures of galaxy
  environment - i. what is `environment'?}, Monthly Notices of the Royal
  Astronomical Society 419~(3) (2011) 2670--2682.
\newblock \href {https://doi.org/10.1111/j.1365-2966.2011.19922.x}
  {\path{doi:10.1111/j.1365-2966.2011.19922.x}}.
\newline\urlprefix\url{https://doi.org/10.11112Fj.1365-2966.2011.19922.x}

\bibitem{Lee:Primack:Behroozi:2016}
C.~T. Lee, J.~R. Primack, P.~Behroozi, A.~Rodríguez-Puebla, D.~Hellinger,
  A.~Dekel, \href{https://doi.org/10.1093/mnras/stw3348}{{Properties of dark
  matter haloes as a function of local environment density}}, Monthly Notices
  of the Royal Astronomical Society 466~(4) (2016) 3834--3858.
\newblock \href
  {http://arxiv.org/abs/https://academic.oup.com/mnras/article-pdf/466/4/3834/10872869/stw3348.pdf}
  {\path{arXiv:https://academic.oup.com/mnras/article-pdf/466/4/3834/10872869/stw3348.pdf}},
  \href {https://doi.org/10.1093/mnras/stw3348}
  {\path{doi:10.1093/mnras/stw3348}}.
\newline\urlprefix\url{https://doi.org/10.1093/mnras/stw3348}

\bibitem{Forero-Romero:Hoffman:Bustamante:2013}
J.~E. {Forero-Romero}, Y.~{Hoffman}, S.~{Bustamante}, S.~{Gottl{\"o}ber},
  G.~{Yepes}, {The Kinematics of the Local Group in a Cosmological Context},
  \apjl 767~(1) (2013) L5.
\newblock \href {http://arxiv.org/abs/1303.2690} {\path{arXiv:1303.2690}},
  \href {https://doi.org/10.1088/2041-8205/767/1/L5}
  {\path{doi:10.1088/2041-8205/767/1/L5}}.

\bibitem{Fattahi:Azadeh:Navarro:2016}
A.~{Fattahi}, J.~F. {Navarro}, T.~{Sawala}, C.~S. {Frenk}, K.~A. {Oman}, R.~A.
  {Crain}, M.~{Furlong}, M.~{Schaller}, J.~{Schaye}, T.~{Theuns}, A.~{Jenkins},
  {The APOSTLE project: Local Group kinematic mass constraints and simulation
  candidate selection}, \mnras 457~(1) (2016) 844--856.
\newblock \href {http://arxiv.org/abs/1507.03643} {\path{arXiv:1507.03643}},
  \href {https://doi.org/10.1093/mnras/stv2970}
  {\path{doi:10.1093/mnras/stv2970}}.

\bibitem{Carlesi:Edorado:Hoffman:2020}
E.~{Carlesi}, Y.~{Hoffman}, S.~{Gottl{\"o}ber}, N.~I. {Libeskind}, A.~{Knebe},
  G.~{Yepes}, S.~V. {Pilipenko}, {On the mass assembly history of the Local
  Group}, \mnras 491~(2) (2020) 1531--1539.
\newblock \href {http://arxiv.org/abs/1910.12865} {\path{arXiv:1910.12865}},
  \href {https://doi.org/10.1093/mnras/stz3089}
  {\path{doi:10.1093/mnras/stz3089}}.

\bibitem{Sorce:Ocvirk:Aubert:2022}
J.~G. {Sorce}, P.~{Ocvirk}, D.~{Aubert}, S.~{Gottloeber}, P.~R. {Shapiro},
  T.~{Dawoodbhoy}, G.~{Yepes}, K.~{Ahn}, I.~T. {Iliev}, J.~S.~W. {Lewis},
  {Reionization time of the Local Group and Local-Group-like halo pairs}, arXiv
  e-prints (2022) arXiv:2207.13102\href {http://arxiv.org/abs/2207.13102}
  {\path{arXiv:2207.13102}}.

\bibitem{Soszynski:Gieren:Pietrzy:2006}
I.~{Soszy{\'n}ski}, W.~{Gieren}, G.~{Pietrzy{\'n}ski}, F.~{Bresolin}, R.~P.
  {Kudritzki}, J.~{Storm}, {The Araucaria Project: Distance to the Local Group
  Galaxy NGC 3109 from Near-Infrared Photometry of Cepheids}, \apj 648~(1)
  (2006) 375--382.
\newblock \href {http://arxiv.org/abs/astro-ph/0605243}
  {\path{arXiv:astro-ph/0605243}}, \href {https://doi.org/10.1086/505789}
  {\path{doi:10.1086/505789}}.

\bibitem{Dalcanton:Williams:Seth:2009}
J.~J. {Dalcanton}, B.~F. {Williams}, A.~C. {Seth}, A.~{Dolphin}, J.~{Holtzman},
  K.~{Rosema}, E.~D. {Skillman}, A.~{Cole}, L.~{Girardi}, S.~M. {Gogarten},
  I.~D. {Karachentsev}, K.~{Olsen}, D.~{Weisz}, C.~{Christensen}, K.~{Freeman},
  K.~{Gilbert}, C.~{Gallart}, J.~{Harris}, P.~{Hodge}, R.~S. {de Jong},
  V.~{Karachentseva}, M.~{Mateo}, P.~B. {Stetson}, M.~{Tavarez}, D.~{Zaritsky},
  F.~{Governato}, T.~{Quinn}, {The ACS Nearby Galaxy Survey Treasury}, \apjs
  183~(1) (2009) 67--108.
\newblock \href {http://arxiv.org/abs/0905.3737} {\path{arXiv:0905.3737}},
  \href {https://doi.org/10.1088/0067-0049/183/1/67}
  {\path{doi:10.1088/0067-0049/183/1/67}}.

\bibitem{Boylan_Kolchin_2013}
M.~Boylan-Kolchin, J.~S. Bullock, S.~T. Sohn, G.~Besla, R.~P. van~der Marel,
  \href{https://doi.org/10.1088/0004-637x/768/2/140}{{THE} {SPACE} {MOTION}
  {OF} {LEO} i: {THE} {MASS} {OF} {THE} {MILKY} {WAY}{\textquotesingle}s {DARK}
  {MATTER} {HALO}}, The Astrophysical Journal 768~(2) (2013) 140.
\newblock \href {https://doi.org/10.1088/0004-637x/768/2/140}
  {\path{doi:10.1088/0004-637x/768/2/140}}.
\newline\urlprefix\url{https://doi.org/10.1088/0004-637x/768/2/140}

\bibitem{Shaya:Tully:2013}
E.~J. Shaya, R.~B. Tully, \href{https://doi.org/10.1093/mnras/stt1714}{{The
  formation of Local Group planes of galaxies}}, Monthly Notices of the Royal
  Astronomical Society 436~(3) (2013) 2096--2119.
\newblock \href
  {http://arxiv.org/abs/https://academic.oup.com/mnras/article-pdf/436/3/2096/13762914/stt1714.pdf}
  {\path{arXiv:https://academic.oup.com/mnras/article-pdf/436/3/2096/13762914/stt1714.pdf}},
  \href {https://doi.org/10.1093/mnras/stt1714}
  {\path{doi:10.1093/mnras/stt1714}}.
\newline\urlprefix\url{https://doi.org/10.1093/mnras/stt1714}

\end{thebibliography}





% \end{thebibliography}
\end{document}